%% file: Final_Version-AIDELMAN.tex
\documentclass{aa}
\usepackage{graphicx}
\usepackage{txfonts}
\usepackage[english]{babel}
\usepackage{color}
\usepackage{natbib}
\usepackage{lscape}
\usepackage{longtable}
\usepackage{setspace}
\usepackage{subfigure}
\usepackage{ulem}

\usepackage{epstopdf}
\begin{document}

\title{BCD spectrophotometry for massive stars in transition phases\thanks{Based on data obtained at Complejo Astronómico El Leoncito, operated under agreement between the Consejo Nacional de Investigaciones Cient\'ificas y T\'ecnicas de la Rep\'ublica Argentina and the National Universities of La Plata, C\'ordoba and San Juan, Argentina, and at Observat\'orio do Pico dos Dias, LNA, Brazil}}
\titlerunning{BCD spectrophotometry for massive stars}

\author{
Y. J. Aidelman\inst{1,2}\fnmsep\thanks{Member of the Carrera del Investigador Cient\'ifico y Tecnol\'ogico, CONICET, Argentina}
\and 
M. Borges Fernandes\inst{3}
\and 
L. S. Cidale\inst{1,2,\star\star}
\and
A. Smith Castelli\inst{1,\star\star}
\and
M. L. Arias\inst{1,2,\star\star}
\and
J. Zorec\inst{4}
\and \\
M. Kraus\inst{5}
\and
A. Torres\inst{1,2,\star\star}
\and
T. B. Souza\inst{3}
\and
Y. R. Cochetti\inst{1,2}\fnmsep\thanks{Fellow of CONICET, Argentina}
}

\authorrunning{Y. Aidelman et al.}

\institute{
Instituto de Astrof\'{\i}sica La Plata, CCT La Plata, CONICET-UNLP, Paseo del Bosque S/N, B1900FWA, La Plata, Argentina.
\and
Departamento de Espectroscop\'{\i}a, Facultad de Ciencias Astron\'omicas y Geof\'{\i}sicas, Universidad Nacional de La Plata (UNLP), Paseo del Bosque S/N, B1900FWA, La Plata, Argentina.
\and
Observat\'orio Nacional, Rua General Jos\'e Cristino 77, 20921-400 S\~{a}o Cristov\~{a}o, Rio de Janeiro, Brazil.
\and
Sorbonne Universit\'e, CNRS, UPMC, UMR7095 Institut d'Astrophysique de Paris, 98bis Bd. Arago, F-75014 Paris, France
\and
Astronomical Institute, Czech Academy of Sciences,
Fri\v{c}ova 298, 251 65 Ond\v{r}ejov, Czech Republic.
}
\offprints{Y. Aidelman \email{aidelman@fcaglp.unlp.edu.ar}}

\date{Received / Accepted}

\abstract
{Stars in transition phases, such as those showing the B[e] phenomenon and luminous blue variables (LBVs), undergo strong, often irregular mass ejection events. The prediction of these phases in stellar evolution models is therefore extremely difficult if not impossible.
As a result, their effective temperatures, their luminosities and even their true nature are not fully known.
}
{A suitable procedure to derive the stellar parameters of these types of objects is to use the BCD spectrophotometric classification system, based on the analysis of the Balmer discontinuity.
The BCD parameters ($\lambda_1$, $D$) have the advantage that they are independent of interstellar extinction and circumstellar contributions.}
{We obtained low-resolution spectra for a sample of 14 stars with the B[e] phenomenon and LBVs. Using the BCD classification system, we derived the stellar and physical parameters. The study was complemented with the information provided by the $JHK$ colour-colour diagram.}
{For each star under investigation, the BCD system gives a complete set of fundamental parameters and related quantities such as luminosity and distance. Among the 14 studied stars, we confirmed the classification of \object{HK\,Ori}, \object{HD\,323771} and \object{HD\,52721} as pre-main sequence Herbig Ae/B[e] stars, \object{AS~202} and \object{HD~85567} as FS~CMa-type stars, and \object{HD\,62623} as sgB[e].
We classified \object{Hen 3$-$847}, \object{CD$-$24\,5721}, and \object{HD~53367} as young B[e] stars or FS~CMa-type candidates, and \object{HD\,58647} as a slightly evolved B[e] star. In addition, \object{Hen\,3$-$1398} is an sgB[e] and \object{MWC~877}, \object{CPD$-$59\,2854} and \object{LHA\,120-S\,65} are LBV candidates.
The stellar parameters of the latter two LBVs are determined for the first time.
We also used the size-luminosity relation to estimate the inner radius of the dust disk around the pre-main and main sequence B[e] stars.}
{Our results emphasise that the BCD system is an important and highly valuable tool to derive stellar parameters and physical properties of B-type stars in transition phases. This method can be combined with near-IR colour-colour diagrams to determine or confirm the evolutionary stage of emission-line stars with dust disks.}


\maketitle

\section{Introduction}
\label{Int}

Over the last few years, there have been huge advances in the description of the internal structure of stars thanks to the inclusion of ingredients like rotation, diffusion, mixing, mass loss, new opacities, diverse metallicities, binarity, etc \citep[c.f][]{Althaus2010, Ekstrom2012, Farrell2019}.
However, although a lot of effort has been made to deepen our understanding of specific short phases in the life cycle of the stars, several of these transition states are still poorly understood. Among them are the stars showing the B[e] phenomenon and LBVs. 
At the moment, the occurrence of these types of objects cannot be predicted with stellar evolution models.
Whether this failure is due to the lack of physical inputs in the stellar evolution models of single stars, or whether this specific evolutionary state might be connected to phenomena occurring during binary star evolution, in particular the occurrence of mergers \citep{Podsiadlowski2006, deMink2013}, is still an ongoing debate. 
Noteworthy in this respect is, however, that post-interactive binary stars display strong similarities to either blue hypergiants or LBVs due to the presence of a dense circumstellar medium similar to that of B[e] stars resulting from a high mass-transfer rate \citep{Farrell2019}.

Stars with the B[e] phenomenon present in their optical spectrum strong Balmer emission lines, as well as forbidden and permitted emission lines of neutral and singly ionized metals.
They also display a strong IR excess and linear polarization due to the presence of a dusty ring-like structure \citep{Coyne1976}.
This group of objects is actually very heterogeneous in an evolutionary sense.
\citet{Lamers1998} have classified many objects exhibiting the B[e] phenomenon into categories such as pre-main sequence intermediate-mass stars (Herbig Ae/B[e], HAeB[e]), evolved low-mass (compact planetary nebulae, cPNB[e]) and high-mass (supergiants, sgB[e]) stars, and also symbiotic systems (symB[e]).
However, more than $50\,\%$ of the Galactic objects with the B[e] phenomenon have an unclear nature, being called unclassified B[e] (unclB[e]) stars, and need to be deeply studied \citep{Condori2019}.
From this last group, \citet{Miroshnichenko2007} identified many stars with properties similar to FS~CMa and proposed a new group: the FS~CMa stars.
These authors propose that they would be B[e] binary systems that are currently undergoing or have recently undergone a phase of rapid mass exchange, showing a spectrum dominated by the B-type star.

The B[e] phenomenon itself seems to be related to the presence of a very complex circumstellar environment around B-type stars.
In general, these stars have dense equatorial disks, as was initially suggested by \citet{Zickgraf1985} and confirmed later with interferometric observations, for instance in \citet{DomicianodeSouza2007, Meilland2010, BorgesFernandes2011, Cidale2012, Varga2019}, among others.
The great challenge in the study of stars with the B[e] phenomenon is to understand how they form their circumstellar envelopes (CE), with very similar spectral characteristics, by going through such different evolutionary stages and, furthermore, how these CEs would subsequently evolve.

The other intriguing group of stars in a transition phase are the LBVs.
These are post-main sequence massive objects characterized by intense and fast mass-loss events, usually called eruptions \citep{Humphreys1994, Mehner2017, Campagnolo2018, Weis2020}.
The origin of these eruptions is not well known, but many studies have indicated an instability region in the upper Hertzsprung–Russell diagram (HRD).
The LBVs located in this region would have high rotational velocities, close to the critical speed \citep{Wolf1989, Clark2005, Groh2009}.
However, \citet{Groh2006} suggested the existence of two groups of LBVs in the Galaxy: one with high and another with low rotational speed.

Since stars with the B[e] phenomenon and LBVs are deeply embedded in their circumstellar media, it is extremely difficult to assign their fundamental parameters and classify them. Quite often, sgB[e] stars are confused with LBVs in quiescence, which share very similar optical spectroscopic characteristics \citep{Kraus2019}.
Moreover, distances to many of the peculiar Galactic objects are not well known either since their luminosity or parallax cannot be properly determined.
Furthermore, low-luminosity Galactic sgB[e] stars often overlap with the most luminous massive pre-main sequence objects (Herbig Ae/B[e]) and cannot easily be disentangled from each other \citep{Kraus2009}.
Therefore, to solve these issues, it is important to obtain precise stellar parameters not only for classification purposes but also as input to models of stellar evolution.
Furthermore, although GAIA parallaxes have been determined for many peculiar Galactic objects, there are still significant differences between photometric and parallax distances for other OB Galactic stars \citep{Shull2019} and, therefore, their luminosities remain poorly estimated.
In addition, stars with the B[e] phenomenon have very extended circumstellar discs or envelopes, whose structure is variable and not necessarily symmetric. However, the radiation they emit can contribute significantly to the overall luminosity of the object, changing thus significantly the position of the overall photo center detected by GAIA, which ends up producing deviations to the parallax estimation \citep{Xu2019, Stassun2021, Chiavassa2022}.

A way to overcome these difficulties is to make use of the empirical BCD spectrophotometric classification system \citep{Barbier1941, Chalonge1952}, based on the analysis of the Balmer discontinuity (hereafter as ``BD'') among F- and B-type stars.
One of the major advantages of this classification system is that the parameters that characterise the BD are independent of the interstellar extinction and the circumstellar contribution \citep[extinction or emission,][]{Zorec1991}.
This system was already successfully applied to study not only stars with the B[e] phenomenon \citep{Cidale2001} but also Be, Bn or B supergiant stars \citep{Zorec2005, Zorec2009, Cochetti2020}.
These authors also illustrated the application of the BCD system to stars showing a second BD (in absorption or emission).
This second component of the BD (SBD) is produced in a low-density gaseous environment of some emission-line stars \citep{Divan1979}. For more details see  \citet[][Sects. \S~3.1 and \S~5.1, respectively]{Aidelman2012,Aidelman2018}.
In this work, we propose to use the BCD spectrophotometric classification system to analyse mainly Galactic stars that either show the B[e] phenomenon or were identified as LBV candidates.
Particularly, the selected stars have an uncertain classification or scarce data available.

The paper has the following structure: observations are described in Sect.~\ref{Obs}; Sect.~\ref{BCD} introduces a summary of the BCD spectrophotometric classification system; Sect.~\ref{Res} gives the obtained stellar parameters and describes the main results of each star; Sect.~\ref{Dis} presents a discussion and our conclusions. In Appendix~\ref{apex}, previous studies on each star are briefly outlined. 

\section{Observations}
\label{Obs}

Low-resolution spectra of a sample of emission-line B-type stars were obtained using a Boller \& Chivens (B\&C) spectrograph attached to the $2.15-$m Jorge Sahade telescope at the Complejo Astron\'omico El Leoncito (CASLEO), San Juan, Argentina.
The spectra were taken on several runs from 2001 to 2012 (see Table~\ref{obslog}).
Two different setups were selected.
Before 2011, we used a grating with $600$~$\ell$\,mm$^{-1}$ ($\#$80, blazed at 4000~\AA), a PM 512 CCD detector, and slit widths of $250~\mu$m and $350~\mu$m (according to the average seeing at the CASLEO).
The covered spectral wavelength interval ranges from $3500$~\AA{} to $4600$~\AA{} with an inverse linear dispersion of $\sim 4.8~\AA${} per two pixels (that is $R \sim 1000$ at $\sim 4500$~\AA).
In 2012, we selected a grating with $600$~$\ell$\,mm$^{-1}$, a CCD detector TEK 1024, and a slit width of $350~\mu$m, covering the spectral range between $3500$~\AA{} and $5000$~\AA.
The inverse linear dispersion is $\sim 5.2$~\AA{} per two pixels ($R \sim 700$ at $\sim 4500$~\AA).

A subset of our program stars was observed with a Boller \& Chivens spectrograph attached to the $1.6-$m Perkin-Elmer telescope at the Laborat\'orio Nacional de Astrof\'isica (LNA), Braz\'opolis, Brazil.
The data were taken on 2012, February 29, and April 11.
The instrumental setup made use of a grating with $600$~$\ell$\,mm$^{-1}$, a slit width of $400~\mu$m, and a Marconi CCD~42-40-1-368 ($2048~\times~2048$ pixels) detector, which provides an inverse linear dispersion of $\sim 2.12$~\AA{} per two pixels ($R \sim 2000$ at $\sim 4500$~\AA).
The wavelength coverage of the spectra is $3500-5000$~\AA.\par
Spectra of He-Ne-Ar (CASLEO) and He-Ar (LNA) comparison lamps were acquired at the same sky position of every science target to perform the wavelength calibration.

All spectra were reduced with the {\sc iraf}\footnote{{\sc iraf} is distributed by the National Optical Astronomy Observatory, which is operated by the Association of Universities for Research in Astronomy (AURA), Inc., under a cooperative agreement with the National Science Foundation} software package, following a standard procedure, such as bias and overscan subtraction, flat-field normalization, and wavelength calibration.
The flux calibration was performed by observing flux standard stars selected from \citet{Hamuy1994} on each night of observation.

From the whole sample of observed emission-line stars, we selected a subset of 14 stars that have a measurable Balmer jump to derive their fundamental parameters and analyse their evolutionary states. This star sample and the details of our observations are given in Table~\ref{obslog}.

\begin{table*}[!ht]
\begin{center}
  \caption{Journal of available spectra listing apparent visual magnitudes (from SIMBAD database), observing dates, the observatory where the spectra were taken, exposure times (the number in brackets gives the total number of exposures).
  The last column lists the different classifications assigned to each star in the literature (for details see the Appendix~\ref{apex} section).}
  \label{obslog}
\tabcolsep 1.0pt

\input{tablas/Table_obslog.tex}

\end{center}
\end{table*}


\section{The BCD spectrophotometric classification system}
\label{BCD}

The BCD spectrophotometric classification system, developed by D. Barbier, D. Chalonge, and L. Divan \citep[called simply as ``BCD system'',][]{Barbier1941, Chalonge1952, Chalonge1973, Chalonge1977}, is a very useful tool for an adequate classification of stars with spectral types between O9 and F9.
It is based on four measurable parameters in the vicinity of the BD: the height of the Balmer jump ($D$), its mean position ($\lambda_1$), and two colour gradients ($\Phi_{uv}$ and $\Phi_b$).

A summary of the BCD parameters is presented here.
The parameter $D$ is defined by,

\begin{equation}
\label{BCD1}
D = \log\left(\frac{F_{+3700}}{F_{-3700}}\right), \\
\end{equation}

\noindent where $F_{+3700}$ is the flux determined by extrapolating the slope of the Paschen continuum at the position $\lambda 3700$, and $F_{-3700}$ is defined by the flux value towards which the highest members of Balmer series converge, given by the intersection of the vertical line at $\lambda 3700$ and the lower curve wrapping the Balmer line cores \citep[see the upper panel of Fig.~3 in][]{Aidelman2012}.
The parameter $\lambda_1$ is defined by the intersection point between a parallel line to the Paschen continuum, traced at $D/2$, and the upper wrapping of the wings of the Balmer lines.
Among dwarf stars, $D$ ranges from near $0.0$~dex to about $0.5$~dex.
The higher values correspond to the A3-4 spectral types, while the lower ones are roughly for O4-5 type stars (the hotter side) or the F9-type stars (the cooler side).
For luminous blue supergiants, the Balmer jump is not seen anymore in the B early spectral types \citep[see Fig.~10 in][]{Zorec2009}.

The colour gradients are determined from observations using the definition provided by \citet{Allen1976},

\begin{equation}
\label{BCD4}
\Phi_{\mathrm{ab}}= - \ln\left[\frac{\lambda_{\mathrm a}^5\, F_{\lambda_{\rm a}}}{\lambda_{\rm b}^{5}\,F_{\lambda_{\rm b}}}\right]/\left(\frac{1}{\lambda_{\rm a}}-\frac{1}{\lambda_{\rm b}}\right),\\
\end{equation}

\noindent where $\lambda_{\rm a}$ and $\lambda_{\rm b}$ are the edges of the selected wavelength ranges. Thus, $\Phi_{\rm uv}$ and $\Phi_{\rm b}$ are determined in the wavelength range of $3150-3700$~\AA{} and $4000-4800$~\AA, respectively.

A detailed description of the BCD system, determination of the BCD parameters, and error estimates can be found in \citet{Zorec2009, Aidelman2012, Aidelman2017}, \citet{Aidelman2018}, and \citet{Zorec2023}.

\begin{figure*}[!ht]
\centering
\includegraphics[width=1.1\hsize,angle=0]{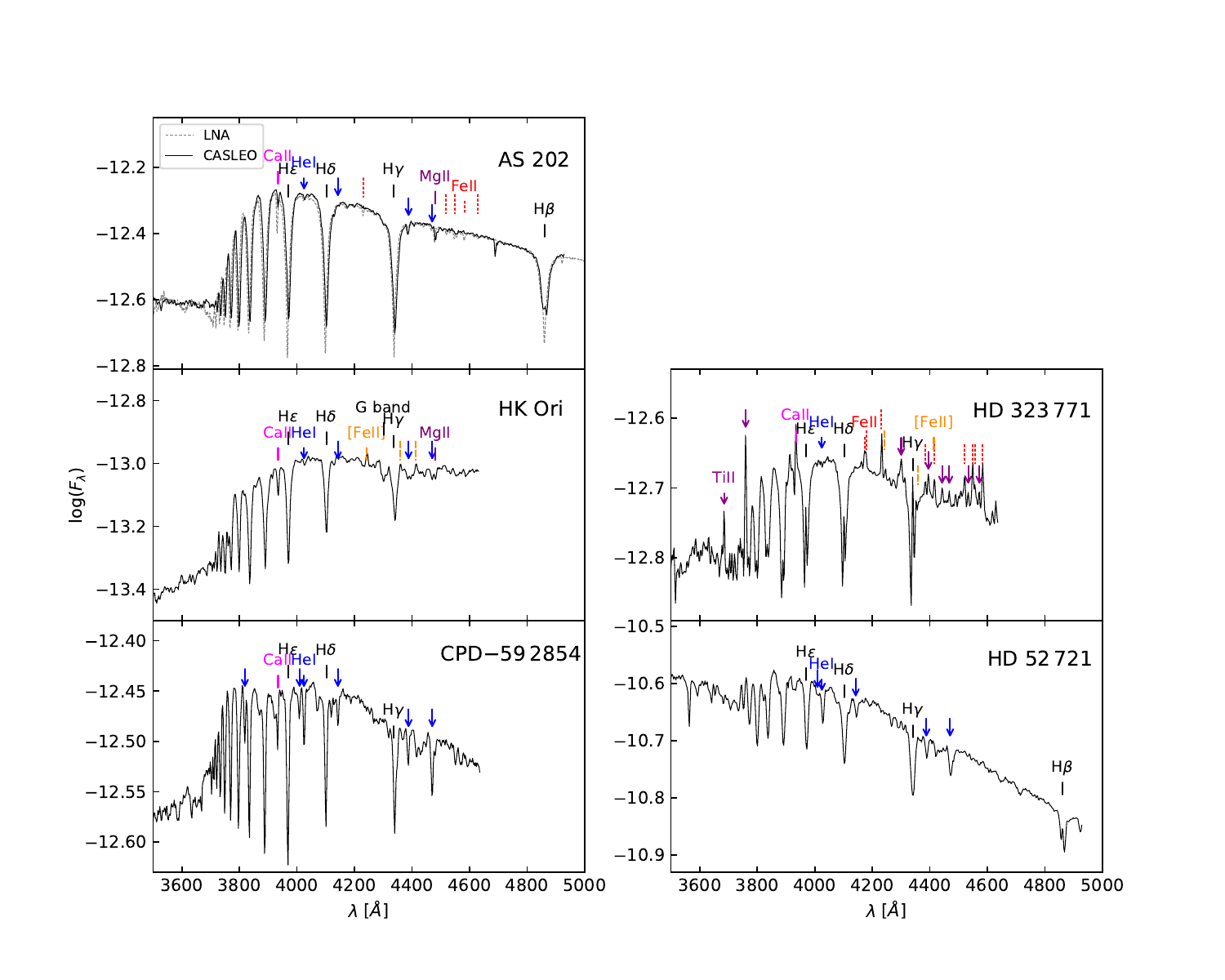}
\caption{Low-resolution flux-calibrated spectra taken in CASLEO and LNA for the sample of B[e] and LBV stars that present the BD, but the SBD is absent. The main spectral lines are highlighted. The flux units are ergs~cm$^{-2}$~sec$^{-1}$~\AA$^{-1}$.}
\label{BD2}
\end{figure*}

\begin{figure*}[!ht]
\centering
\includegraphics[width=1.1\hsize,height=1.22\hsize,angle=0]{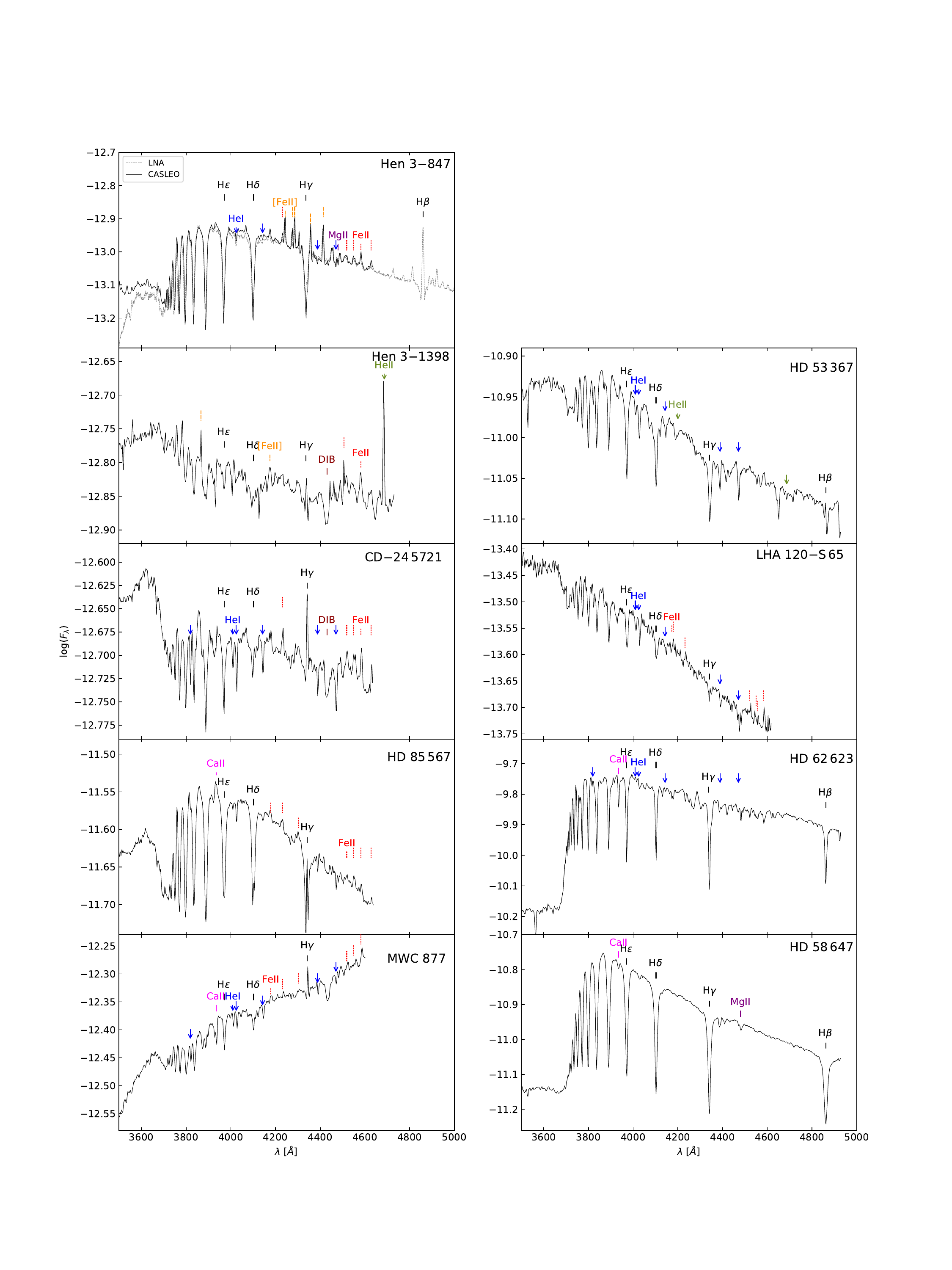}
\caption{Low-resolution flux-calibrated spectra taken in CASLEO and LNA for the sample of B[e] and LBV stars that present the BD and the SBD. The main spectral lines are highlighted. The flux units are in ergs~cm$^{-2}$~sec$^{-1}$~\AA$^{-1}$.
}
\label{BD1}
\end{figure*}

The height and position of the main BD are related to the physical properties of the stellar photospheres and allow us to derive the effective temperature ($T_{\rm eff}$), surface gravity ($\log\,g$), and absolute visual ($M_{\rm v}$) and bolometric ($M_{\rm bol}$) magnitudes through empirical relations obtained by \citet{Zorec2009, Zorec1991, Zorec1986}.
Furthermore, since there is also a direct relationship between the BCD and the MK systems, it is also possible to derive the spectral type and luminosity class.

The slope of the Paschen continuum gives the extinction, $A_{\rm v} = R_{\rm v}\,E(B-V)$.
For this purpose, we prefer using the following expression for $E(B-V)$ given by \citet{Aidelman2012},

\begin{equation}
  \label{BCD6}
  \begin{gathered}
 E(B-V)=0.68\,(\Phi_{\mathrm{b}}-\Phi_{\mathrm{b}}^{0})=0.75\,(\Phi_{\mathrm{bb}}-\Phi_{\mathrm{bb}}^{0})
 \;{\textrm{mag}}\\
  A_{\rm v} = 2.11\,(\Phi_{\mathrm{b}}-\Phi_{\mathrm{b}}^{0})=2.33\,(\Phi_{\mathrm{bb}}-\Phi_{\mathrm{bb}}^{0})
 \;{\textrm{mag}},\\
 \end{gathered}
\end{equation}

\noindent where $\Phi_{\mathrm{bb}}$ and $\Phi_{\mathrm{bb}}^{0}$ are the observed and intrinsic colour gradients in the wavelength range $4000-4600$~\AA, while $\Phi_{\mathrm{b}}$ and $\Phi_{\mathrm{b}}^{0}$ are the observed and intrinsic colour gradients for the interval $4000-4800$~\AA.
The $E(B-V)$ expression was obtained assuming that $R_{\rm v} = 3.1$, which is the usual value for the diffuse interstellar medium \citep{Cardelli1989, Schlafly2011}.

To perform all these measurements, we used the interactive code MIDE3700\footnote{The code MIDE3700 is available upon request.}, written by Y.~A. in Python language \citep{Aidelman2017}.
By fitting the Balmer and Paschen continua and the upper and lower envelopes of the Balmer lines, the code computes both the BCD and fundamental stellar parameters.

\section{Results}
\label{Res}

Figures~\ref{BD2} and \ref{BD1} display the flux-calibrated spectra of the 14 stars that show a measurable Balmer jump. Typical lines identified in the B-type spectra are:
\ion{H}{i}, 
\ion{He}{i},
\ion{Mg}{ii},
\ion{Ca}{ii},
\ion{Si}{ii},
\ion{Si}{iii},
\ion{Si}{iv},
\ion{C}{ii}, and
\ion{He}{ii} (for the hottest spectral types).
Some other distinctive emission lines of \ion{Fe}{ii}, [\ion{Fe}{ii}], \ion{O}{ii}, \ion{Ti}{ii}, etc. are also indicated.
The obtained BCD parameters from the observed BD are given in Table~\ref{BCDpar}.

Only five stars show the typical photospheric BD: \object{AS\,202}, \object{HK\,Ori}, \object{HD\,52721}, \object{HD\,323771}, and \object{CPD\,$-$59\,2854} (see Fig.~\ref{BD2}).
The peculiar star in the LMC (\object{LHA\,120$-$S\,65}) shows an SBD in emission, like the other six Galactic stars: \object{Hen\,3$-$847}, \object{Hen\,3$-$1398} (very weak), \object{CD$-$24\,5721},
\object{HD\,53367}, \object{HD\,85567}, and \object{MWC\,877}.
By contrast, in \object{HD\,62623} and \object{HD\,58647} the SBD is in absorption (see Fig.~\ref{BD1}).

Table~\ref{ParFun} lists the stellar parameters: spectral type ($ST$), $T_{\rm{eff}}$, $\log\,g$, $\Phi^{0}$ ($\Phi_{\rm b}^{0}$ or $\Phi_{\rm bb}^{0}$ as appropriate), $M_{\mathrm{v}}$, and $M_{\mathrm{bol}}$, which were derived from the pair ($\lambda_1, D$); and the total absorption $A_{\rm V}$, which was computed with Eq.~\ref{BCD6}.
The luminosity classes for stars with $\lambda_{1} - 3700 > 70~\AA$ are outside of the BCD spectral-type calibration and are indicated as "V:" in Table~\ref{ParFun}.
On the other hand, as the BCD calibration for $\log\,g$ of early-type B supergiants has not been done yet; these values were estimated from the $T_{\rm{eff}}-\log\,g$ relationship found by \citet[][given in their Fig.~7]{Haucke2018}.

An additional set of parameters is given in Table~\ref{ParFun2}, such as the stellar luminosity in terms of the solar luminosity, $\log~(L/L_\sun)$;  the inverse of the ``flux-weighted gravity'', $\log~(\mathcal{L}/\mathcal{L}_{\sun})$,  where $\mathcal{L}=T_{\rm eff}^{4}/g$ \citep{Langer2014}; the true distance modulus, $(m - M_{\mathrm{v}})_0$, and the corresponding distance, $d$.
Column 6 in Table~\ref{ParFun2} gives the distance, $d_{G}$, to each star that results from the measured parallaxes by the Gaia mission, Early Data Release 3 \citep[EDR3,][]{Bailer-Jones2021}, and the corresponding errors.
Values of $\log (L/L_\sun)$ were derived from $M_{\rm bol}$.

\subsection{The HR Diagram}

\begin{figure*}[t!]
  \resizebox{\hsize}{!}{\includegraphics{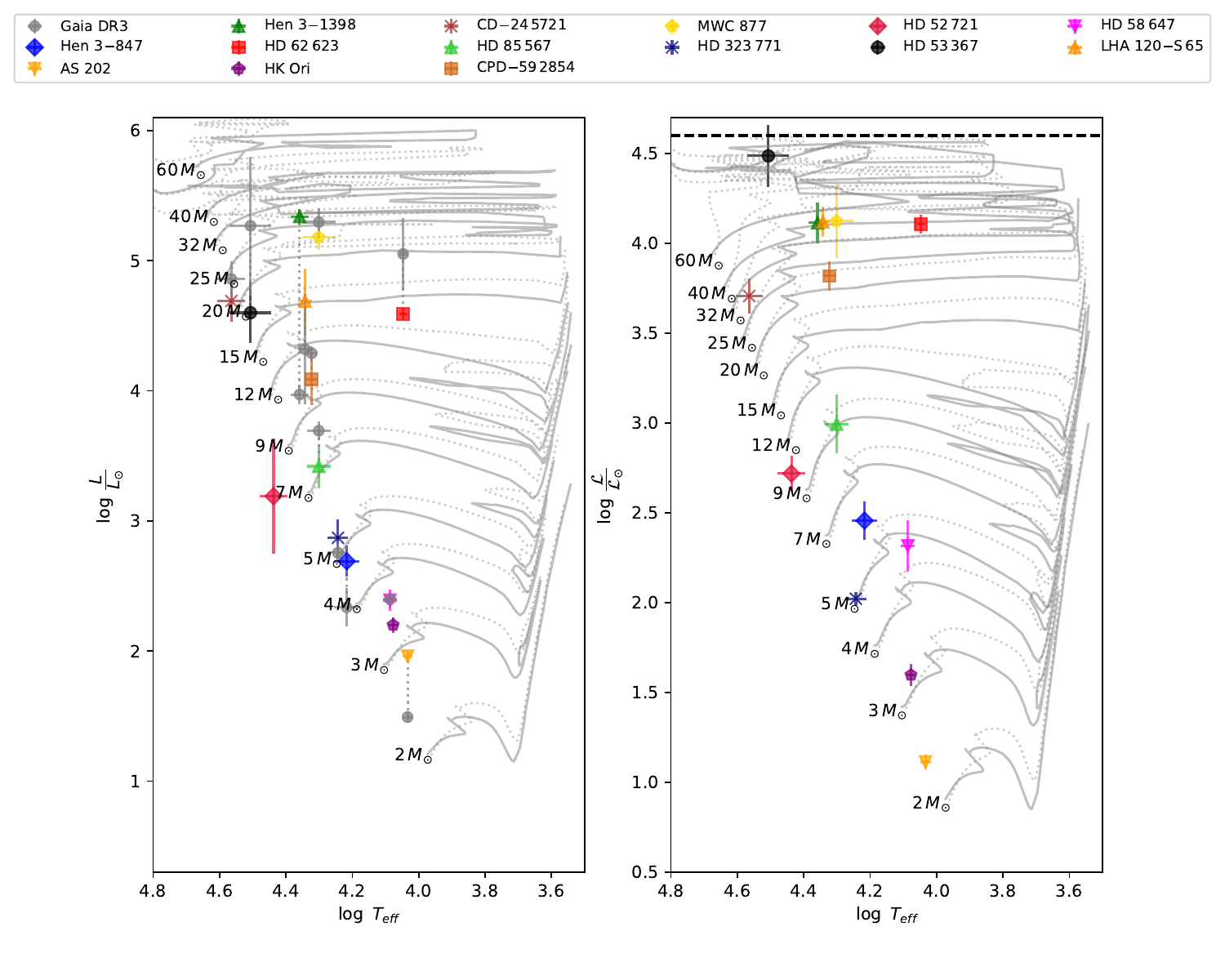}}
  \caption{{\em Left:} The HR diagram for the stars of our sample (coloured symbols). The grey circles indicate the positions of stars as if they were located at distances given by Gaia EDR3 \citep{Bailer-Jones2021}. The BCD and Gaia positions of the same object are connected with dotted lines. Evolutionary tracks with and without rotation ($\Omega/\Omega_{c} = 0.4$ and $\Omega/\Omega_{c} = 0.0$ in dotted and solid lines, respectively) were taken from \citet{Ekstrom2012}. The positions of \object{Hen\,3$-$847} and \object{AS\,202} correspond to mean measurement values obtained at CASLEO and LNA.
  {\em Right:} The sHR diagram. The dashed black line at $\log \mathcal{L}/\mathcal{L}_{\odot} = 4.6$ corresponds to the Eddington luminosity limit for a hydrogen-rich composition \citep{Langer2014}.
}
\label{sHRD}
\end{figure*}

To analyse the results, the locations of the studied stars were plotted in the HRD (see Fig.~\ref{sHRD}, left panel) together with the evolutionary tracks computed with and without rotation ($\Omega/\Omega_{c} = 0.4$ and $\Omega/\Omega_{c} = 0.0$), respectively by \citet{Ekstrom2012}.
These values are compared with the luminosity expected from  Gaia's distances ($d_{G}$, grey symbols), which was obtained using the apparent magnitude $m_{\rm V}$ from SIMBAD, the bolometric correction given by \citet{Flower1996} and the colour excess $E(B-V)$ derived from the BCD system.
We also constructed the spectroscopic HR diagram \citep[sHRD,][]{Langer2014} that can be seen in Fig.~\ref{sHRD}, right panel.
This diagram provides a distance-independent measurement of the stellar luminosity-to-mass ratio. This kind of diagram could help to discuss the evolutionary stage of a star, particularly when it is ambiguous.

\begin{table}[ht]
\begin{center}
\caption{BCD parameters of our star sample. 
The SBD can be in emission (E), absorption (A), or absent (N) is also indicated.
\label{BCDpar}}
\tabcolsep 2 pt

\input{tablas/Table_BCDpar.tex}

\end{center}
\end{table}

In the HRD, we found that most of the stars of our sample are in the pre- or main-sequence stage, while only a few are evolved objects. Instead, in the sHRD, many stars appear to be more massive and evolved than in the classical HRD.  
In both diagrams of Fig.~\ref{sHRD}, stars \object{HD\,52721}, \object{HD\,53367}, \object{HD\,323\,771}, \object{Hen\,3$-$847}, \object{CD$-$24\,5721}, and \object{AS\,202} are located on the Zero Age Main Sequence (ZAMS) or very near to it. The known mass discrepancy problem is briefly addressed in Sect.~\ref{evolution}.

\begin{table*}
\begin{center}
\caption{Stellar parameters obtained from the BCD system.\label{ParFun}}

\input{tablas/Table_ParFun.tex}

\end{center}
\end{table*}

\begin{table*}
\begin{center}
\caption{Additional set of stellar parameters for the star sample. \label{ParFun2}}
\tabcolsep 2.5 pt

\input{tablas/Table_ParFun2.tex}

\end{center}
\end{table*}

\subsection{The $JHK$ colour-colour diagram}

\begin{figure*}[!ht]
\centering
\resizebox{\hsize}{!}{\includegraphics{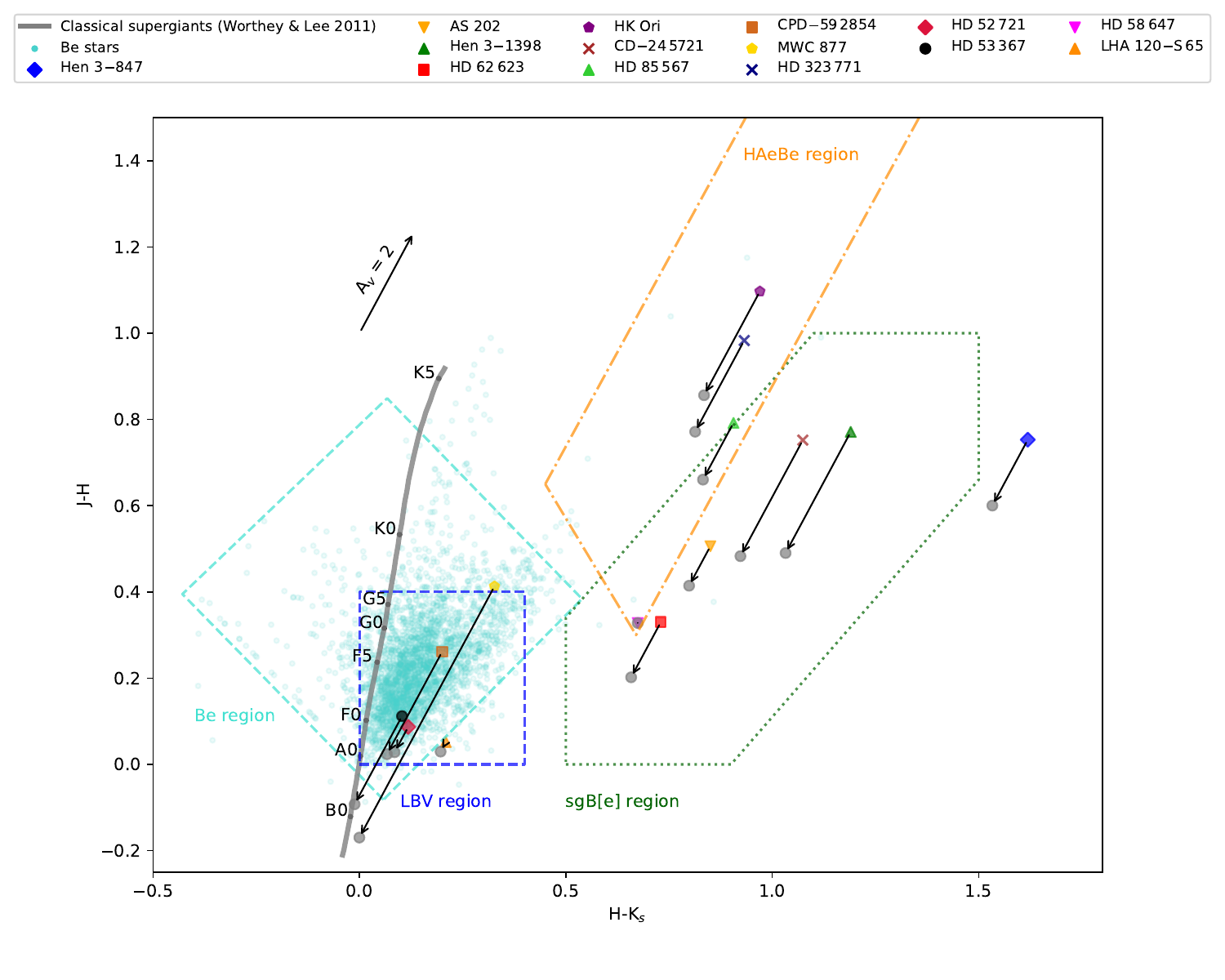}}
\caption{Near-infrared colour-colour diagram. Apparent colours are marked by coloured symbols and true colours are represented by filled grey circles. Data of the same stars are connected with arrows.}
\label{JHKD}
\end{figure*}

We use the NIR $JHK$ colour-colour diagram (shown in Fig.~\ref{JHKD}) as a supplementary tool to discuss the controversial nature of the stars in our sample.
There, we point out particular regions where B[e] supergiants (green box), LBVs (blue box), and Herbig Be/Ae stars with a moderate or large IR excess (orange box) are usually located \citep{Cochetti2020b, Kraus2019, Oksala2013}.
The thick grey line, on the left, defines the location of classical supergiants \citep{Worthey2011}.
Data of the same stars are connected with arrows.
The NIR colours were corrected using the $A_{\rm V}$ values derived from the BCD spectrophotometric system given in Table \ref{ParFun2} and the reddening law was obtained from \citet{Wang2014}.

For completeness, we also defined in the NIR colour-colour diagram the region where stars that present the Be phenomenon are located (box and symbols in cyan).
We used 3374 Be stars taken from the literature\footnote{The Be Star Spectra database (BeSS); the CoRoT spectroscopic Be stars Atlas; SIMBAD database; \citet{Cochetti2020, Hou2016, Jaschek1982, McSwain2005, McSwain2009, Raddi2015, Vioque2020}, and \citet{Witham2008}.} to delimit this region.
This region is contaminated with Herbig Ae/Be stars that have a small NIR excess \citep{Aidelman2023}.

Based on the $JHK$ true colours of the stars we identify four stars (\object{Hen\,3$-$1398}, \object{HD\,62623}, \object{CD-24\,5721} and \object{AS\,202}) in the region of the B[e] stars \citep[mainly supergiants,][]{Oksala2013}; three stars (\object{HD\,52721}, \object{CPD-59\,2854}, and \object{LHA\,120$-$S\,65}) are inside the LBV box; four stars (\object{HD\,58647}, \object{HD\,85567}, \object{HD\,323\,771}, and \object{HK\,Ori}) are among the Herbig Ae/Be stars, although \object{HD\,58647} and \object{HD\,85567} are located in a region that is also shared by the B[e] supergiants.
Finally, we identify three stars (\object{Hen\,3$-$847}, \object{HD\,53367} and \object{MWC\,877}) that are outside of the previously defined regions.

\subsection{Comments on individual objects}

In previous sections, for each star of our sample, we analysed their low-resolution spectra, making line identification and obtaining the BCD parameters.
Below we comment on our main results regarding the nature and stellar parameters of the studied stars. See Appendix A for details on previous studies.

\bigskip
\noindent {\bf \object{Hen\,3$-$847}} (\object{CD$-$48\,7859}, \object{V1028\,Cen}, \object{IRAS\,12584$-$4837}): 
This star has an uncertain evolutionary stage. 
Our low-resolution spectra show an SBD in emission revealing the presence of a dense circumstellar envelope (see Fig. \ref{BD1}).
The highest members of the Balmer series are in absorption.
The H$\gamma$ line shows an incipient emission at its core, while the H$\beta$ line (shown in the LNA spectrum) displays a broad photospheric absorption with a narrow central emission line.
Emission lines of \ion{Fe}{ii} and [\ion{Fe}{ii}] are also present.
From the spectra taken at different epochs, we derive similar BCD ($\lambda_{1}$, $D$)  parameters, although the colour gradients are slightly dissimilar because they are measured on spectra with different wavelength ranges.
The mean stellar parameters are: $T_{\rm eff}= 15\,020$~K and  $\log\,g= 3.7$ (see Table \ref{BCD}).
These values are in excellent agreement with those found in the literature (see Table~\ref{StellarParamLit}).

We estimated a mean visual absolute magnitude $M_{\mathrm v} = -1.13$~mag, $A_{\mathrm v} = 1.21$~mag and a distance $d \sim 1210 \pm 180$~pc.
This value is in between the distances given by \citet[][$d=1660$~pc]{Verhoeff2012} and \citet[][$d=997_{-218}^{+379}$~pc]{Bailer-Jones2018}, but has a large discrepancy with the new value listed in Gaia DR3 ($649_{-63}^{+93}$~pc). 

Based on the location of the star in the HR and sHR diagrams (shown in Fig.~\ref{sHRD}), the star could be either a main sequence or pre-ZAMS star  with an initial stellar mass of $5$~M$_\odot$ or $6$~M$_\odot$.
However, the star is located outside of the Herbig Ae/Be and sgB[e] regions in the NIR colour-colour diagram.
Although the anomalous colours of this star resemble those observed in the FS~CMa type objects  \citep{Miroshnichenko2007}, the star has not been confirmed as a binary system yet.

\bigskip
\noindent {\bf  \object{AS\,202}} (\object{Hen\,3$-$174}, \object{IRAS\,08307$-$3748}, \object{FX\,Vel}):
The low-resolution spectra display mainly broad photospheric H lines in absorption that reveal an A0V star.
We also identify weak \ion{He}{i} lines.

The H and \ion{Ca}{ii} lines are variable. Particularly, the H$\beta$ line profile presents a narrow and variable emission component at its core.
This component seems to be weaker than the one reported by \citet{Miroshnichenko2007b}.
No other emission lines are observed in the optical spectral range shown in Fig.~\ref{BD1}.

Using the BCD spectrophotometric system, we classify the primary star as an A0-1~V spectral type located at a mean distance of $470$~pc.
The obtained spectral type agrees with the classification done by \citet{Eggen1978} and \citet{Aldoretta2015}, while the distance derived from the BCD parameters is somewhat greater than the one obtained from $Gaia$ mission \citep[$d=353_{-6}^{+6}$~pc and $d=335_{-4}^{+3}$~pc,][respectively]{Bailer-Jones2018, Bailer-Jones2021}.

Based on the location of \object{AS\,202} in the HR diagram, it can be either a main sequence or a pre-main sequence star with an initial stellar mass of $2.3-3$~M$_\odot$ (see Fig.~\ref{sHRD}).

\bigskip
\noindent {\bf \object{Hen\,3$-$1398}} (\object{CPD$-$38\,6814}, \object{MWC\,878}, \object{IRAS\,17213$-$3841}, \object{CD$-$38\,11837}):
The stellar spectrum shows the high members of Balmer lines in absorption.
The core of H$\gamma$, \ion{He}{ii}~$\lambda 4686$ and lines of \ion{Fe}{ii} and [\ion{Fe}{ii}] are in emission.
Strong DIBs at $\lambda4430$~\AA\, and $\lambda4502$ ~\AA\, are present.

Based on our BCD results, the presence of a small BD ($D = 0.02$), and the location of the star inside the B[e] supergiant region in the NIR colour-colour diagram, we suggest that \object{Hen\,3$-$1398} is a massive B0-B1\,Ia supergiant.
The obtained $T_{\rm eff} = 25\,000$~K and $E(B-V) = 0.77$~mag correspond to extrapolated values from the BCD calibration curves. They agree with those calculated by \citet[][25\,400~K and 0.74~mag]{Carmona2010}.
However, the spectral type of this star could be slightly earlier, i.e., O9~I, considering the presence of a very strong emission in the \ion{He}{ii} $\lambda4686$ line and uncertainties arising from the extrapolation.
This line is a luminosity indicator in O-type stars \citep{Conti1974, Walborn1977, Sota2014}.
\citet{Venero2002} showed that \ion{He}{ii} line profiles observed among OB stars can be attributed to different conditions in the stellar wind and their intensities are very sensitive to the gravity effect.
From a theoretical point of view, the \ion{He}{ii} $\lambda4686$ line can be observed in emission also in B supergiant stars with $T_{\rm eff} \sim 25\,000$~K.

The BCD absolute magnitude calibration for hot supergiants is not completed (for $\lambda_{1} - 3700 < 30~\AA$), and extrapolating values for small $\lambda_1$ will introduce large errors. To estimate a rough distance to \object{Hen\,3$-$1398} we adopted the $M_{\rm v}$ value from \citet{Cox2000}.
In this way, taking $M_{\rm v} = -6.43$~mag, $m_{\rm v} = 10.62$~mag, and $E(B-V)=0.77$~mag, the distance modulus is $14.53$~mag that leads to a distance of $8536\pm236$~pc.
Moreover, the presence of a strong DIB $\lambda4430$ can be consistent with a high interstellar reddening, and thus, with a possible large distance.

\bigskip
\noindent {\bf  \object{HD\,62623}} (\object{3\,Puppis}, \object{HR\,2996}, \object{IRAS\,07418$-$2850}):
The star exhibits a spectrum dominated by H lines in absorption and has an important SBD in absorption (see Fig.~\ref{BD1}) that reveals the presence of a prominent and dense circumstellar gaseous disc. 

From the BCD system, we found that \object{HD\,62623} is a B9\,Ib supergiant star \citep[in concordance with][]{Stephenson1971} with a large colour excess $E(B-V)= 0.35\pm0.06$~mag. 
Then, assuming $M_{\rm v} = -6.25$~mag \citep[value taken from][]{Cox2000}, for $R_{\rm v} = 3.1$, we calculate a distance modulus of $9.11$~mag.
This value gives  a distance of $662\pm54$~pc, which is in excellent agreement with the distance obtained by \citet[][$d=650\pm100$~pc]{Meilland2010} and  \citet[][$d=664^{+240}_{-140}$~pc]{Bailer-Jones2018} but it disagrees with the Gaia EDR3 estimate ($d = 1124^{+357}_{-198}$~pc).

The NIR colours of the star support its nature as a B[e] supergiant in the $JHK$ colour-colour diagram.

\bigskip
\noindent {\bf  \object{HK\,Ori}} (\object{MWC\,497}, \object{IRAS\,05286$+$1207}, \object{TYC\,709$-$857$-$1}):
The spectrum shown in Fig.~\ref{BD1} reveals weak lines of \ion{He}{i} and \ion{Mg}{ii} consistent with a B8V spectral type.
The H lines are observed in absorption, but H$\gamma$ and H$\delta$ are weak and might be partially filled in by emission.
The [\ion{Fe}{ii}] lines are observed in emission. The G band is also present, supporting the presence of a cool secondary companion.
  
From the BCD classification system, we derived a spectral type B9V and a $T_{\rm{eff}}= 11\,940$~K, which are consistent with the \ion{He}{i} features identified in the spectrum.
These BCD stellar parameters agree with the spectral type and temperature obtained by \citet{Whittet1983} and \citet{Garrison1978}, respectively.
  
The NIR $JHK$ colours put the star in the Herbig Ae/Be region.
Moreover, the locus of the star in the HR diagram is compatible with a pre-main/main sequence star of $3-3.3$~M$_\odot$ at a distance of $630$~pc.
The location on the sHR diagram supports this result.
The distance and stellar mass are somewhat greater than the values found in the literature ($\sim 460$~pc, $1.7-2$~M$_\odot$, see Tables~\ref{StellarParamLit} and \ref{StellarParamLit2}).

\bigskip
\noindent {\bf  \object{CD$-$24\,5721}} (\object{Hen\,3$-$52}, \object{ALS\,612}, \object{IRAS\,07370$-$2438}):
The observed B\&C spectrum displays \ion{He}{i} lines in absorption whose intensities are consistent with an MK spectral type O8.
The SBD is in strong emission, suggesting the star has a dense circumstellar envelope (see Fig.~\ref{BD1}).

Using the BCD spectrophotometric system, we derived a $T_{\rm eff} = 36\,600$~K.
We also estimated an initial stellar mass of $21$~M$_\odot$ and a distance of $4\,200$~pc.
Our spectral classification (O8~IV) agrees with that done by \citet{Nordstrom1975} but it does not match the previous BCD classification done by \citet{Cidale2001} since it was based on measurements performed in a non-flux calibrated spectrum.
  
We found that this object is located near the ZAMS (see Fig.~\ref{sHRD}).
However, it presents NIR colours expected for sgB[e]s, as shown in Fig.~\ref{JHKD}.
We return to the discussion on its nature in Sect.~\ref{Dis}.

\bigskip
\noindent {\bf  \object{HD\,85567}} (\object{CPD$-$60\,1510}, \object{Hen\,3$-$331}, \object{V\,596\,Car}):
Our spectrum shows an SBD in emission, revealing the presence of a dense circumstellar envelope.
The star presents broad H absorption lines with a narrow central emission.
The \ion{He}{i} lines are also in absorption, while the lines of \ion{Fe}{ii} and \ion{Ca}{ii} are in emission. 

The BCD system yields a B3~III star with a $T_{\rm eff}=20\,000$~K, a value that is in very good agreement with that derived by \citet{Miroshnichenko2001} and \citet{Manoj2006}.
The star has NIR $JHK$ colours {\bf as typically} expected for Herbig Ae/Be stars and sgB[e] stars. 
This ambiguity is also evident in Fig.~\ref{sHRD}, where the object seems to be in the HRD as a main sequence star, while in the sHRD, it appears as an evolved star.
We obtained a distance to the star of 913~pc, in excellent agreement with the measurement obtained from Gaia mission \citep[][$d=1002_{-28}^{+30}$~pc and $d=1036_{-15}^{+17}$~pc, respectively]{Bailer-Jones2018, Bailer-Jones2021} and by \citet[][$d=910$~pc]{Fairlamb2015}.

\bigskip
\noindent {\bf \object{CPD$-$59\,2854}} ( \object{WRAY\,15$-$689}, \object{IRAS\,10538$-$5958}):
The B\&C spectrum around the BD exhibits intense hydrogen, helium, and \ion{O}{ii} absorption lines.
Apart from the fact that the H$\gamma$ line seems to be filled in by emission, we do not observe any other emission feature in the observed spectral wavelength range.

Using the BCD system, the stellar parameters of the star have been derived for the first time (see Tables~\ref{ParFun} and \ref{ParFun2}).
We have assigned the star a B2~II spectral type, close to the classification by \citet{Graham1970}.
In addition, the distance determined in this work ($d = 3.505 \pm 0.768$~kpc) agrees very well with that estimated by \citet[][$d=3.825^{+0.480}_{-0.386}$~kpc and $3.144_{-0.108}^{+0.104}$~kpc, respectively]{Bailer-Jones2018, Bailer-Jones2021}. 
According to the star's locus in the HRD and sHRD, the initial stellar mass might be $9-10$~M$_\sun$ or $20-25$~M$_\sun$, respectively.
This last value is consistent with the fact that the star is inside the LBV region in the CCD (see Fig.~\ref{JHKD}).

\bigskip
\noindent {\bf  \object{MWC\,877}} (\object{HD\,323154}, \object{CD$-$38\,11806}, \object{Hen\,3$-$1393}, \object{IRAS\,17197$-$3901}):
The blue region of the spectrum shows many narrow absorption lines of hydrogen and helium,  while the core of H$\gamma$ and the \ion{Fe}{ii} forest lines are seen in emission (see Fig.~\ref{BD1}).
The height of the Balmer jump is consistent with a spectral type B3~Ia in excellent agreement with that reported by \citet{Vijapurkar1993}.
This star presents an SBD in emission and seems to be located in a region with a very anomalous interstellar extinction, $A_{\rm V}= 4.96$~mag derived by the BCD.
Consequently, the corrected $JHK$ colours put the star outside (but nearby) the LBV region.

The stellar distance estimated through the BCD method ($1.297\pm0.046$~kpc) is in excellent agreement with that derived from GAIA parallax \citep[][$d=1.270_{-0.120}^{+0.147}$~kpc]{Bailer-Jones2018} but lower than that reported from GAIA Early Data Release 3 \citep[EDR3, $d=1.501_{-0.056}^{+0.069}$~kpc,][]{Bailer-Jones2021}. 
 
The presence of an SBD and the location of the star in the NIR colour-colour and HR diagrams point to a supergiant star with a disk. We do not discard that the star could be an LBV candidate.

\bigskip
\noindent {\bf \object{HD\,323771}} (\object{CD$-$39\,11602}, \object{Hen\,3$-$1425}, \object{IRAS\,17306$-$3921}):
Our spectrum shows strong and broad \ion{H}{} lines in absorption. 
In addition, these lines exhibit narrow emission components in their cores that are seen up to the end of the Balmer series.
The star also presents very weak \ion{He}{i} lines. The \ion{Ca}{ii}, \ion{Fe}{ii}, \ion{Ti}{ii} and [\ion{Fe}{ii}] lines are in emission.

The BCD spectral type, B5 V:, is in very good agreement with that reported by \citet{Vieira2003} and \citet{Herbst1975}. Although, the latter proposed that the star is B5 or later.
We derived a distance of $1\,432\pm129$~pc which is greater than but still in fair agreement with the distance derived from GAIA DR2 and EDR3 \citep[][$d=1072^{+69}_{-61}$~pc and $d=1074_{-29}^{+26}$~pc, respectively]{Bailer-Jones2018, Bailer-Jones2021}.

Based on the features in the stellar spectrum, the BCD parameters, and the position of the star in the HR and $JHK$ diagrams, we classify it as a Herbig Ae/B[e] star of $\sim 5.5$~M$_{\odot}$.

\bigskip
\noindent {\bf \object{HD\,52721}} (\object{GU\,CMa}, \object{MWC\,164}, \object{IRAS\,06594$-$1113}):
The optical spectrum exhibits broad lines of H and \ion{He}{i}.
The H$\beta$ line presents a weak emission in the core. Very tiny emission lines of the \ion{Fe}{ii} forest are present.
The BCD parameters suggest a spectral type B2~V: with $T_{\rm eff} = 27\,361 \pm 2611$:~K and $\log\,g =4.4$ at a distance $d = 479 \pm 76$~pc.
This distance was obtained considering that the star is a visual binary with a magnitude difference of $\sim 0.7$~mag \citep{Aitken1932}.
The effective temperature is higher than the values found in the literature, but the distance agrees with that reported by \citet{Fairlamb2015}.

We also found that the star has anomalous $JHK$ colours, which resemble those typically seen in LBV and Be stars (see Fig.~\ref{JHKD}).
Its location in the HR diagram indicates that it is a young stellar object of $7$~M$_\sun$.
This value is lower than that obtained by \citet[][$9.5$~M$_\sun$]{Fairlamb2015}, which better agrees with the star's locus in the sHRD.

\bigskip
\noindent {\bf \object{HD\,53367}} (\object{V\,750\,Mon}, \object{MWC\,166}, \object{IRAS\,07020$-$1022}):
The spectrum of this star displays \ion{He}{i} and \ion{He}{ii} lines.
All the lines are in absorption with the exception of the presence of a weak and narrow emission in the H$\beta$ core.
The SBD is in emission.
The star is associated with a dark nebula but has a surface gravity more representative of evolved stars.

We derived a B0~III spectral type for the star.
This very luminous and massive B[e] star is also located within the star-forming CMa~OB1 association at a distance of $d=1131$~pc.
This value is lower than the measurement provided by Gaia EDR3 ($d = 1599^{+926}_{-458}$~pc), which is given with high uncertainty.

\bigskip
\noindent {\bf \object{HD\,58647}} (\object{BD$-$13\,2008}, \object{IRAS\,07236$-$1404}): 
The star shows a spectrum with the Balmer lines and the \ion{Mg}{ii} $\lambda4481$ line in absorption.
The star has a spectral type B9~IV and presents an SBD in absorption.
This SBD could be related to the presence of the detected magnetosphere \citep{Jarvinen2019}.

The BCD spectral type accords well with that given by \citet{Jaschek1980}.
The distance obtained ($295\pm7.7$~pc) is also in good agreement with the value reported by \citet{vandenAncker1998} and Gaia EDR3 ($280^{+80}_{-50}$~pc and $302\pm2$~pc, respectively).

The star is located within the zero-age main sequence and has $JHK$ colours consistent with those of Herbig~Ae/Be or sgB[e] stars.
The star also presents polycyclic aromatic hydrocarbons (PAH) bands \citep{Acke2010}.

\bigskip
\noindent {\bf \object{LHA\,120$-$S\,65}} (\object{AL\,18}, \object{2MASS\,J04470544$-$6816167}): 
The spectrum of this star is shown for the first time.
In the observed spectral region, it displays the \ion{He}{i} lines and the highest members of the Balmer series in absorption.
A few \ion{Fe}{ii} lines of permitted transitions are seen in emission (see Fig.~\ref{BD1}).
The Balmer jump corresponds to a B2~Ib star, {\bf showing} an SBD in emission.
The derived BCD stellar parameters are $T_{\rm eff}=22\,000$~K and $\log\,g=2.6$. Based on the star's locus in the HR and $JHK$ diagrams, it could be an LBV in a quiescence phase. 

The distance modulus derived here matches well with the weighted mean distance modulus of the LMC, $18.477 \pm 0.004$~mag ($d = 49.6$~kpc) \citep{Pietrzynski2019}. 
Whilst the value reported in the GAIA DR3 release disagrees completely with the expected for the LMC.

\section{Discussion}
\label{Dis}

We studied low-resolution flux-calibrated spectra for a sample of 14 stars in transition phases showing the B[e] phenomenon and LBVs.
These stars are usually deeply embedded in their circumstellar environments rendering it difficult to derive proper stellar parameters and disentangle their evolutionary stage.

In this work, we used the BCD spectrophotometric classification system, which is independent of interstellar extinction and circumstellar contributions, and derived the stellar parameters.
In addition, based on the star's loci in the HRD, sHRD, and the $JHK$ colour-colour diagram, we discuss their evolutionary stages.

\subsection{Fundamental parameters}
\label{Fun_Par}

In general, the BCD system has proved to be a reliable tool to derive the stellar parameters and physical properties of normal B-type stars and those with emission lines, for example, Be, B[e] and LBV stars \citep[][among others]{Cidale2001, Zorec2005, Aidelman2018}.
However, for stars showing the B[e] phenomenon, we cannot be completely sure that the Balmer discontinuity corresponds only to some photospheric-like regions of the central star. 
In the same way, the spectroscopic lines of these complicated objects originate from both the stellar photosphere and circumstellar regions. Therefore, both techniques might not accurately reflect the intrinsic properties of the central object.
Nevertheless, the Balmer discontinuity is often less contaminated by circumstellar features than the optical part of the spectrum. Thus, the derived fundamental parameters may likely better characterise the properties of the deep photospheric layers.

In the cases of interacting binary systems, the determination of the intrinsic properties of the stars is more complex.
For example, a change of the spectral type from the blue to the red part of the spectrum might occur, as was reported by \citet{Strom1983} in \object{HK\,Ori}.
This could also be the case of \object{HD\,62623}.
For this star, the BCD spectral type and effective temperature (B9~Ib and $11\,136 \pm 318$~K) are higher than the value obtained by \citet[][A3~I, $T_{\rm eff} = 8\,500$~K]{Miroshnichenko2020} who also derived the binary orbital parameters.
The discrepancy might come from measurements made in different spectral ranges, epochs and orbital phases. 
For instance, in the blue part of the spectrum (Fig.~\ref{BD2}),
the intensity of the \ion{Ca}{ii}~K line relative to the hydrogen lines (in particular to H$\epsilon$) are consistent with a B9~I spectral type \citep{Gray2009}.
In addition, our spectrum exhibits weak lines of \ion{He}{i} at $\lambda$~4026, 4143, and 4471~\AA.
However, the optical lines in the spectral range $4400 - 4600$~\AA{} resemble an A3 I star \citep{Miroshnichenko2020}.
To discuss the discrepancy found, we also analysed our spectrum with the ULySS software package\footnote{\href{http://ulyss.univ-lyon1.fr/}{http://ulyss.univ-lyon1.fr/}} \citep[University of Lyon Spectroscopic analysis Software,][]{Koleva2009}, and obtained $T_{\rm eff} = 10\,392 \pm 179$~K and $\log\,g = 1.89 \pm 0.05$.
These new values are also consistent with a spectral type B9 I \citep{Cox2000}.
We believe that the spectral type of the star depends on the spectral region analysed and that its spectrum also varies with the orbital phase.

Something similar could occur in the binary \object{HD\,85567} where the intensity of the Balmer jump leads to a $T_{\rm eff} = 20\,000 \pm 1639$~K while \citet{Khokhlov2017} obtained $\sim 15\,000 \pm 500$~K from the optical range. However, in both works, the derived luminosity for the star is the same.

Both \object{HD\,62623} and \object{HD\,85567} are good candidates to study in more detail their spectral variability against the orbital phase using the BCD classification system. This would allow us to discuss the influence of binarity in the determination of the fundamental parameters with the BCD method or the possible constraints on its use.

\subsection{On the evolutionary stage}
\label{evolution}

\begin{figure}[t!] 
\resizebox{\hsize}{!}{\includegraphics{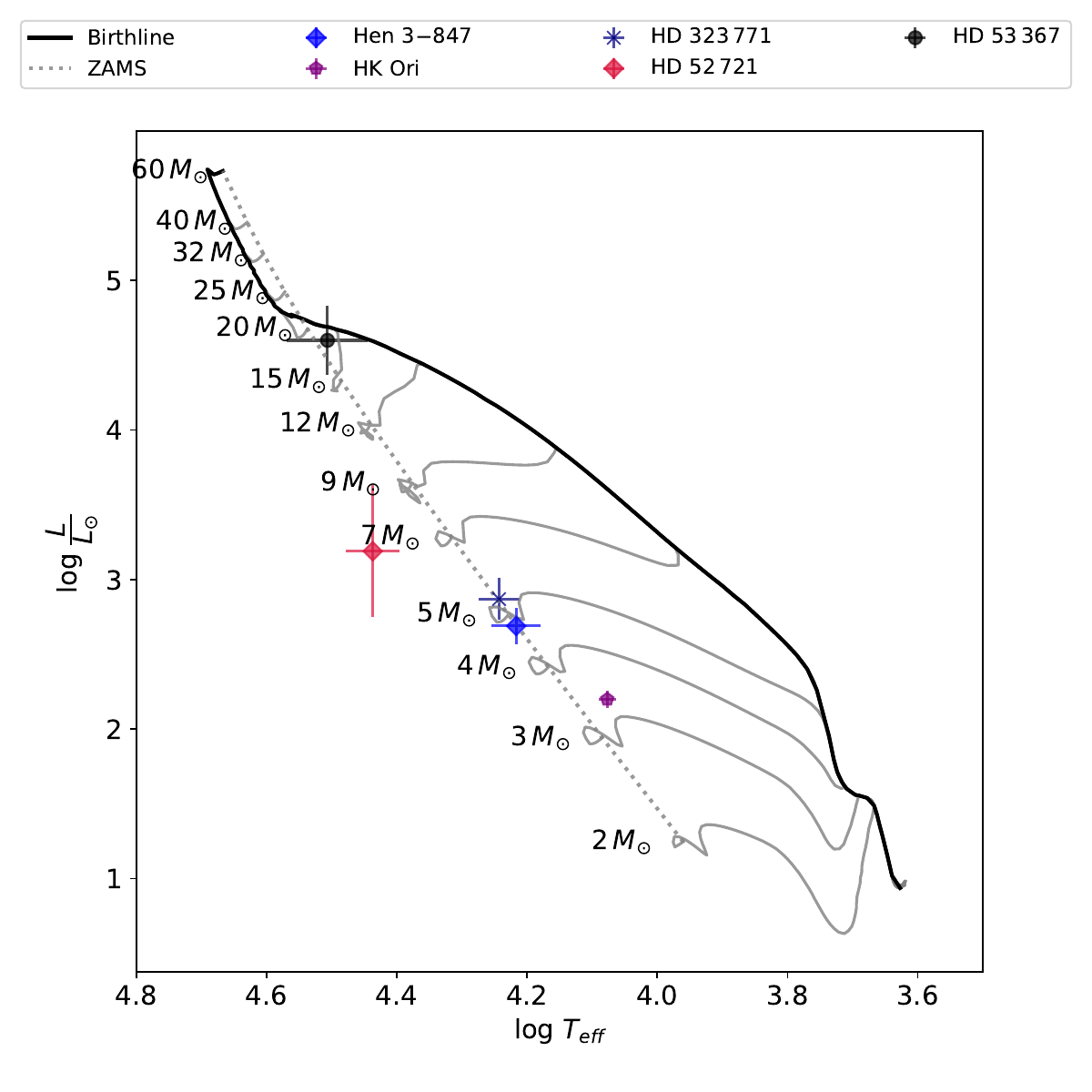}}
\caption{Location of the young stellar objects in the HR diagram. Pre-main sequence evolutionary tracks are from \citet{Haemmerle2019}. The zero-age main sequence (dotted line) and birth line (solid line) are also plotted. Numbers refer to the stellar masses (in solar masses) when reaching the zero-age main sequence.}
\label{Track2}
\end{figure}

\begin{itemize}
    \item[a)] Pre- and main-sequence stars:

    In a recent work, \citet{Aidelman2023} confirmed three groups of HAeBe stars that present different amounts of NIR radiation excess using different reddening-free $Q$ parameters. 
    A group of HAeBes with small IR excess (Group I) shares properties with LBVs and some CBe stars.
    These stars are located in the region bounded by $-0.2 < (H-Ks) < 0.2$ and $-0.2 < (J-H) < 0.2$ \citep{Chen2016} and host huge amounts of cool dust at large distances from the star \citep{Oksala2013}.
    A second group with moderate IR excess (Group II) is located in the region shown in Fig.~\ref{JHKD} as HAeBe. 
    Their excesses of IR radiation probably indicate the influence of a bright nebula associated with low temperatures or thermal emissions from the surrounding dust~\citep{Hillenbrand1992}.
    Group III with a large IR excess and hot dust \citep{Oksala2013} shows similar properties to some sgB[e]s stars.

    From the present study, we found nine stars located on the ZAMS or main-sequence (MS) in the HRD region: \object{Hen\,3$-$847}, \object{HK\,Ori}, \object{AS\,202}, \object{CD$-$24\,5721}, \object{HD\,52721}, \object{HD\,53367}, \object{HD\,58647},  \object{HD\,85567}, and \object{HD\,323771}.
    Some of these stars are associated with star-forming regions.

    Considering the locations of \object{HK\,Ori} and \object{HD\,323771} in the HRD and NIR colour-colour diagram, they are Herbig Ae/Be stars with moderate IR excess.
    This classification agrees with previous ones found in the literature (see Appendix A).
    In Table~\ref{Object_type}, we defined them as Group~II HAeB[e] objects according to their NIR colours.
    
    \object{HD\,52721} and \object{HD\,53367} have NIR colours typically seen among LBVs or CBe stars, as shown in Fig.~\ref{JHKD}. 
    Both locations also coincide with the region delimited for the Group I HAeBe stars with less or no infrared excess.
    Since the HAeBes have masses between $\sim 2$~M$_{\odot}$ and $\sim 15$~M$_{\odot}$ \citep{Herbig1960, Hillenbrand1992, Vioque2022}, only \object{HD\,52721} fulfills the HAeBe classification criterion.
    Instead, \object{HD\,53367}, which also belongs to a multiple system \citep{Corporon1999, Pogodin2006}, has a mass slightly higher than $15$~M$_{\odot}$ (Figs.~\ref{sHRD} and \ref{Track2}), and the BCD classification system points it out as a B0 III-type star. 
    Therefore, we conclude it is a young massive B[e] star, likely of the FS CMa-type.

    Regarding \object{HD\,58647} and \object{HD\,85567}, they are located on the main sequence, as shown in Fig.~\ref{sHRD} but they are not immersed in star-forming regions. 
    On the other hand, their NIR colours locate them in the area shared by HAeBe and sgB[e] objects, suggesting processed circumstellar dust.
    It is also essential to address that the derived BCD distances are in very good agreement with the GAIA EDR3 ones indicating that the spectral type and luminosity class derived in this work are plausible.
    
    The last two stars, \object{AS\,202} and \object{CD$-$24\,5721}, are reported as binaries or multiple systems (see Appendix~\ref{apex}) and are neither associated with star-forming regions.
    Both stars, like \object{HD\,85567}, are main-sequence objects showing strong IR emission due to the presence of processed dust as they share photometric properties with the sgB[e].
    The dust emission could originate from mass exchange binary interactions, as typically seen in the FS~CMa objects.
    Furthermore, considering the location of \object{CD$-$24\,5721} in the HRD, which is near the ZAMS, it is a massive star that might undergo rejuvenation.\\

    \item[b)] Evolved stars:

    According to the BCD classification, we found five stars (\object{Hen\,3$-$1398}, \object{HD\,62623}, \object{CPD$-$59\,2854}, \object{MWC\,877}, and \object{LHA\,120$-$S\,65}) that evolved from the main sequence and are located in the supergiant region, see Fig.~\ref{sHRD}.
    Two of them, \object{HD\,62623} and \object{Hen\,3$-$1398}, also lie in the sgB[e] region in the colour-colour diagram, therefore we accept them as sgB[e].
    The other three stars (\object{CPD$-$59\,2854}, \object{LHA\,120$-$S\,65}, and \object{MWC\,877}) were classified as LBV candidates, according to the BCD spectral types and NIR colours.
    
\end{itemize}

Finally, Table~\ref{Object_type} summarises the proposed evolutionary classification scheme for our studied star sample.
The young stellar objects are plotted in Fig.~\ref{Track2}, where the pre-main sequence evolutionary tracks are from \citet{Haemmerle2019}.

\begin{table}
\caption{Evolutionary stage determined in this work for the stars in our sample.}\label{Object_type}
\begin{center}
\begin{tabular}{llr}
\hline
\hline
\noalign{\smallskip}
~~~~ID     & ST$_{BCD}$ & Object type\\
\hline \noalign{\smallskip}
\object{Hen\,3$-$847}     & B6 IV &  young B[e] / FS CMa\tablefootmark{c}\\
\object{AS\,202}          & A1 V: & FS CMa\\
\object{Hen\,3$-$1398}    & B1 Ia & sgB[e]\\
\object{HD\,62623}        & B9 Ib & sgB[e]\\
\object{HK\,Ori}          & B9 V & group II HAeB[e]\\
\object{CD$-$24\,5721}    & O8 IV & young B[e] / FS CMa\tablefootmark{c}\\
\object{HD\,85567}        & B3 III & FS CMa\\
\object{CPD$-$59\,2854}   & B2 II & LBV\tablefootmark{c}\\
\object{MWC\,877}         & B3 Ia & LBV\tablefootmark{c}\\
\object{HD\,323\,771}     & B5 V: & group II HAeB[e]\\
\object{HD\,52721}        & B2 V: & group I HAeB[e]\\
\object{HD\,53367}        & B0 III & young B[e] / FS CMa\tablefootmark{c}\\
\object{HD\,58647}        & B9 IV & Slightly evolved B[e]\\
\object{LHA\,120$-$S\,65} & B2 Ib & LBV\tablefootmark{c}\\
\hline
\end{tabular}
\tablefoot{
  \tablefoottext{c}{Star classified as a candidate to FS CMa or LBV type.}
 }
\end{center}
\end{table}

\subsection{HRD vs sHRD}
    The initial masses derived from the HRD and the pre-main sequence evolutionary tracks are very similar. In addition, when comparing the locations of our stars in the HR with the sHR diagram, we find that stars with initial masses lower than around $10$ M$_\odot$ lead to similar masses in both diagrams (at least the observed discrepancy is of the order of $1$~M$_{\odot}$). However, the mass discrepancy becomes significant for stars with masses above $10$~M$_{\sun}$.

    A plausible explanation for this discrepancy might be attributed to binarity.
    This could be the case of \object{HD\,53367}, reported as a multiple system (see Fig.~\ref{sHRD}).
    Supporting this idea, \citet{Burkholder1997} found a good agreement between evolutionary and dynamical masses up to $25$ M$_\sun$. They also concluded that in semidetached or contact binary systems, the derived masses for each component are significantly less than that expected from single-star evolutionary models, suggesting that mass lost from each star is generally not accreted onto the companion.
    However, the same authors pointed out that for larger masses, HRD or sHRD would have limited applicability because of the lack of precise distance and inclination determinations, or because the systems fill (or overfill) their Roche lobes, and would be valid only for the study of single stars.
    Therefore, overall, for low- and intermediate-mass stars, one can use the spectroscopically determined effective temperature and gravity to obtain the stellar mass-to-luminosity ratio, with both HRD and sHRD being adequate \citep{Langer2014}.

    For the case of our LBV candidates and sgB[e] stars, the HR diagram indicates a mass clearly below the value given in the sHR diagram.
    According to \citet{Langer2014}, both diagrams can be used together to identify overluminous stars. These authors explained that the discrepancy could be due to an additional mass-loss process not accounted for in the single-star evolutionary tracks or an unusual mixing process inside the star.
    \citet{Castro2018} also observed mass discrepancies in their analysis of massive stars in the Small Magellanic Cloud. They found that the stellar masses derived from the sHRD for Be and most luminous stars are systematically larger than those obtained from the conventional HR diagram. These authors suggested that the well-known spectroscopic mass-discrepancy problem may be linked to the fact that both groups of stars have outer envelopes that are nearly gravitationally unbound. When considering the previous discussion, the initial masses related to the position in the sHRD of our LBV candidates (\object{CPD$-$59\,2854}, \object{MWC\,877} and \object{LHA\,120$-$S\,65}) seems to be more adequate compared with the typical values given by \citet{Smith2017}.

\subsection{The size-luminosity relation}

Interferometric studies have revealed that for most Herbig Ae/Be stars, there is a correlation between the dust disk inner size ($r_{\rm in}$) and the stellar luminosity \citep[cf.][]{Marcos-Arenal2021}. 
There are two main models that propose a linear relationship between $\log r_{\rm in}$ and $\log\, (L/L_{\sun})$.
The magnetospheric accretion scenario \citep[“optically thin-MA model”,][]{Uchida1985, Shu1994} suggests that the stellar magnetic field connects the optically-thin gas disks with the central star.
In this model, gaseous discs are inner to dust discs.
In this scenario, the correlation ($r_{\rm in} \propto L_\star^{1/2}$) was probed through the spatially resolved NIR continuum emission \citep{Monnier2005, Ababakr2016}.
Although the found correlation has a significant scatter, the characteristic size of a given source seems to be directly related to the radius of dust sublimation of relatively large grains ($\sim 1~\mu$m), with sublimation temperatures in the range $1\,000-2\,000$ K.

The second model consists of an optically thick gas disk that partially shields the stellar radiation, allowing the dust to survive closer to the star, hence shrinking the value of $r_{\rm in}$.
This gaseous disk could reach the stellar surface and directly accrete through a hot, dense boundary layer. This is called the ``optically thick-BL model" \citep{Hillenbrand1992}.

In this subsection, we estimate and discuss the dust disk inner sizes ($r_{\rm in}$)  of our star sample classified as HAeB[e], young massive stars and main-sequence stars (Table~\ref{Object_type}) from the BCD stellar luminosity.
Table~\ref{rin} lists the $r_{\rm in}$ values for the upper and lower limits using thin-MA and thick-BL models \citep[see Fig. 3 and formulae in][]{Marcos-Arenal2021}.
Mean values are given in brackets.

Mainly, \object{HD\,85567} was observed with AMBER/VLTI by \citet{Vural2014}.
These authors modelled the observations with geometric ring models with inner radii approximately 3–5 times smaller than that predicted from an optically thin disk model.
They remarked that their result further supports the existence of an optically thick gaseous inner disk shielding the stellar radiation.
Therefore, using the size-luminosity correlation for the optically thick-BL model, we estimated a mean $r_{\rm in}~\sim~0.95$~AU which fits within the interval $0.8-1.6$~AU obtained with the AMBER instrument.

\object{HD\,58647} is another object with continuum K-band interferometry listed in \citet{Marcos-Arenal2021}'s work.
The BCD luminosity for this star ($\log (L/L_{\odot}) = 2.39$) predicts for the optically thick-BL model a mean value of $\log r_{\rm in} = -0.3$, which matches quite good with the small measured $\log\,r_{\rm in}=-0.29$ reported in that work.

\object{HD\,52721} is one of the nearest objects of our sample.
According to the size-luminosity relation, the expected mean inner radius of its dust disk is $r_{\rm in} \sim 0.76$~AU or $r_{\rm in}~\sim 5.26$~AU for optically thick-BL and thin-MA models, respectively.
Together with \object{HD\,323771} ($r_{\rm in}~\sim 0.55$~AU for a thick-BL model), both stars are good candidates to study their circumstellar disks with optical interferometry.

Finally, it is interesting to note that if we assume that the size-luminosity relation for HAe/Be stars also holds for sgB[e]s, we obtained for \object{HD\,62623} a $r_{\rm in}$ between $1.6$~AU (lower limit) and $4.2$~AU (upper limit) for the optically thick-BL model.
This upper limit estimate is compatible with the gaseous disk radius ($4.7-5$~AU) derived by \citet{Maravelias2018} by fitting forbidden optical emission lines and NIR CO emission bands using a Keplerian rotating disk model \citep{Kraus2007, Kraus2000}. 
These $r_{\rm in}$ values are also comparable with that estimated by \citet[][$r_{\rm in} = 15-20$~R$_{\star}$]{Stee2004}.
Thus, the size-luminosity relation should be explored in the context of evolved B[e] stars with interferometric measurements, for which the dusty disks are formed most likely from equatorially outflowing dense wind material \citep[e.g.][]{BorgesFernandes2007}.

\begin{table*}
\caption{Estimations of inner radii of dust disks for main-sequence and HAeB[e] stars using the optically thin-MA and thick-BL models. The values correspond to lower and upper limits for each model. The mean values of each interval are in brackets.\label{rin}}
\begin{center}
\begin{tabular}{lclc}
\hline
\hline
\noalign{\smallskip}
~~~~ID     &      $\log (L/L_{\sun})$ &    ~~~~~~~~~$r_{\rm in}$ &    $r_{\rm in}$\\
           &   & ~~~~~Thin-MA & Thick-BL \\ 
           &   & ~~~~~~~~[AU] & [AU] \\
\hline \noalign{\smallskip}
\object{Hen\,3$-$847}$^{LNA}$      & $2.69\pm0.13$ & $0.43-5.48\, (2.95)$ & $0.26-0.67\, (0.46)$\\
\object{Hen\,3$-$847}              & $2.76\pm0.11$ & $0.47-5.94\, (3.20)$ & $0.28-0.71\, (0.50)$\\
\object{AS\,202}$^{LNA}$           & $1.94\pm0.11$ & $0.18-2.31\, (1.25)$ & $0.13-0.32\, (0.22)$\\
\object{AS\,202}                   & $1.76\pm0.11$ & $0.15-1.88\, (1.01)$ & $0.11-0.27\, (0.19)$\\
\object{HK\,Ori}                   & $2.20\pm0.06$ & $0.25-3.12\, (1.68)$ & $0.16-0.41\, (0.29)$\\
\object{CD$-24\,5721$}             & $4.69\pm0.16$ & $4.33-54.78\, (29.55)$ & $1.86-4.70\, (3.28)$\\
\object{HD\,85567}                 & $3.42\pm0.17$ & $1.00-12.70\, (6.85)$ & $0.54-1.36\, (0.95)$\\
\object{HD\,323\,771}              & $2.87\pm0.14$ & $0.53-6.74\, (3.64)$ & $0.31-0.79\, (0.55)$\\
\object{HD\,52721}                 & $3.19\pm0.44$ & $0.77-9.74 (5.26)$ & $0.43-1.08\, (0.76)$\\
\object{HD\,53367}                 & $4.60\pm0.23$ & $3.90-49.36\, (26.64)$ & $1.71-4.30\, (3.00)$\\
\object{HD\,58647}                 & $2.39\pm0.08$ & $0.31-3.88\, (2.09)$ & $0.20-0.50\, (0.35)$\\
\hline
\end{tabular}
\end{center}
\end{table*}


\section{Conclusions}
 
The BCD system has proved to be an easy-to-use tool to derive the star's stellar parameters and physical properties with the B[e] phenomenon and LBVs.
This information was supplemented with the NIR colour-colour diagram, which is especially useful for revealing stars' nature in transition phases.
We derived spectral types and luminosity classes of all stars in the sample.

Among the 14 studied stars, we confirmed the classification of \object{HK\,Ori}, \object{HD\,323771} and \object{HD\,52721} as pre-main sequence Herbig Ae/B[e] stars (the first two show moderate NIR radiation excess while the latter shows only a tiny amount), \object{AS~202} and \object{HD~85567} as FS~CMa-type stars, and \object{HD\,62623} as sgB[e].
We classified \object{Hen 3$-$847}, \object{CD$-$24\,5721}, and \object{HD~53367} as young B[e] stars or FS~CMa-type candidates, and \object{HD\,58647} as a slightly evolved B[e] star. In addition, \object{Hen\,3$-$1398} is an sgB[e] and \object{MWC~877}, \object{CPD$-$59\,2854} and \object{LHA\,120-S\,65} are LBV candidates.
The stellar parameters of these two latter LBVs have been obtained for the first time.

In addition, using the size-luminosity relation, we estimated the inner radii expected for the dust disks around the young objects and provided suitable candidates for spectro-interferometric observations.


\begin{acknowledgements}
This research has made use of a) the SIMBAD database, operated at CDS, Strasbourg, France, operated at the Department of Theoretical Physics and Astrophysics of the Masaryk University, and the NASA Astrophysics Data System (ADS); b) data from the European Space Agency (ESA) mission {\it Gaia} (\url{https://www.cosmos.esa.int/gaia}), processed by the {\it Gaia} Data Processing and Analysis Consortium (DPAC, \url{https://www.cosmos.esa.int/web/gaia/dpac/consortium}), and BeSS database (operated at LESIA, Observatoire de Meudon, France: \url{http://basebe.obspm.fr}). Funding for the DPAC has been provided by national institutions, in particular, the institutions participating in the {\it Gaia} Multilateral Agreement.
  
YJA, MLA and AFT thank the financial support from CONICET (PIP 1337), and the Universidad Nacional de La Plata (Programa de Incentivos 11/G160), Argentina. MBF acknowledges financial support from  the National Council for
Scientific and Technological Development – CNPq - Brazil (grant number: 307711/2022-6).
MK acknowledges financial support from the Czech Science Foundation (GA\,\v{C}R 20-00150S). The Astronomical Institute Ond\v{r}ejov is supported by RVO:67985815.
YRC acknowledges support from a CONICET fellowship.

This project has received funding from the European Union's Framework Programme for Research and Innovation Horizon 2020 (2014-2020) under the Marie Skłodowska-Curie Grant Agreement No. 823734.
 
\end{acknowledgements}

\bibliographystyle{aa}
\bibliography{referencias.bib}
\newpage
\appendix

\section{}
\label{apex}

Below we provide a brief description,  found in the literature, of the objects of our sample. Previous reported stellar parameters are summarized in Tables~\ref{StellarParamLit} and \ref{StellarParamLit2}.

\bigskip
\noindent {\bf \object{Hen\,3$-$847}}: 
This star was included in the list of southern emission-line stars of \citet{Wray1966} and \citet{Henize1976}.
It was considered as a pre-main sequence Herbig Ae/Be star by \citet{The1994, deWinter2001, Valenti2003}, and more recently by \citet{Verhoeff2012} and \citet{Fairlamb2015}.
Based respectively on IRAS and IUE data, \citet{Parthasarathy1993} and \citet{Gauba2003} considered it as a post-AGB candidate, although the latter did not rule out the possibility that the star can be a massive young OB supergiant.
The spectral data and the spectral energy distribution (SED) of \object{Hen\,3$-$847} indicate that it has a high effective temperature and presents both warm and cold circumstellar dust \citep{Gauba2003, Gauba2004}.
\citet{Lagadec2011} recognized this star as a proto-planetary nebula and showed that its circumstellar medium cannot be resolved by VLT/VISIR images.
The star was classified as an unclB[e] star by \citet{Lamers1998}.

The stellar mass and radius were determined by \citet[][$6.2\pm1.2$~M$_{\odot}$, $5.9\pm0.6$~R$_{\odot}$]{Verhoeff2012} and \citet[][$4.7$~M$_{\odot}$, $4.5$~R$_{\odot}$]{Fairlamb2015} who, in addition, estimated an age for the star of $0.59$~Myr.
In more recent work, \citet{Arun2019} derived a stellar mass of $2.93$~M$_{\odot}$ and an age of $2.28$~Myr making use of the $Gaia$ parallax. 

\bigskip
\noindent {\bf \object{AS\,202}}:
It was cited as an object with a bright H$\alpha$ emission \citep{Merrill1950} and classified as an {\it ``early-type emission line star with a large far-IR excess''} \citep{The1994}.
\citet{deWinter2001} listed it as a possible Herbig Ae/Be star.
The star exhibits optical photometric variability \citep{Strohmeier1968} and is reported as an eclipsing binary of spectral type B9III-IV with a period of $1.0526$~days \citep{Eggen1978, Malkov2006}, but \citet{tisserand2013} assigned this star a spectral type of A3III.
\citet{Miroshnichenko2007b} proposed \object{AS\,202} as an object with the B[e] phenomenon (based on the presence of  [\ion{O}{i}] emission lines) and classified it as a FS~CMa binary system with components of spectral type A+K.
Recently, \citet{Condori2019} not only derived the stellar parameters but also found indications for a binary system, with a low-mass star and a cool companion, associated with a circumbinary disk or ring.

\bigskip
\noindent {\bf \object{Hen\,3$-$1398}}:
This star is listed in the catalogues of \citet{Merrill1949} and \citet{Henize1976}.
It was classified as a nova \citep{Henize1961} when the star reached its maximum brightness in 1949.
Observations made by \citet{Guglielmo1993} show the presence of a large NIR excess due to warm dust, which contains a weak silicate $10$~$\mu$m emission feature \citep{Volk1991}.
\citet{The1994} classified the object as a B[e] star and \citet{Miroshnichenko2001} suggested that it would be a hot and luminous main-sequence O9/B0 star with warm dust.
            
Based on new high-resolution spectra, \citet{Carmona2010} provided new estimations for the stellar parameters (see Tables~\ref{StellarParamLit} and \ref{StellarParamLit2}) and suggested that it would be an isolated pre-main sequence Herbig Ae/Be star.
These authors also reported Balmer emission lines (H$\alpha$, H$\beta$, and H$\delta$) and double-peaked broad \ion{He}{i} and \ion{He}{ii} emission lines.
Narrow forbidden emission lines of [\ion{O}{i}], [\ion{O}{iii}], [\ion{N}{ii}], and [\ion{S}{iii}] dominate the optical spectrum.

The star has a low projected rotational velocity \citep[$v \sin i = 50$~km\,s$^{-1}$,][]{Miroshnichenko2001}, a mass of $20\pm4$~M$_{\odot}$ and a radius of $8.9\pm0.8$~R$_{\odot}$ \citep{Verhoeff2012}.

\bigskip
\noindent {\bf \object{HD\,62623}}:
This object was classified in the Yale Bright Stars Catalogue \citep{Hoffleit1982} as A2\,Iabe star.
A detailed optical analysis of its spectrum was done by \citet{Chentsov2010} who derived a spectral type of A2.7$\pm$0.3~Ib, being up to now, the unique sgA[e] in our Galaxy \citep{Meilland2010}.
The binary nature was reported for the first time by \citet{Johnson1946} and confirmed later by several authors \citep{Iriarte1965, Levato1972, Hinkle1987, Rovero1994, Plets1995, Miroshnichenko2007, Millour2011}.
According to \citet{Miroshnichenko2020}, this binary system evolved from a $6.0 + 3.6$~M$_{\odot}$ pair with an initial orbital period of $\sim 5$~days, experienced an almost conservative mass transfer and is currently observed as a $0.8 + 8.8$~M$_{\odot}$ object with the B[e] phenomenon.
The current derived orbital period is $137.4 \pm 0.1$ days.

Interferometric observations of \object{HD\,62623} revealed a gaseous and dusty circumstellar environment \citep{Bittar2001, Stee2004}.
The dusty and gaseous emission components were disentangled by modelling VLTI/MIDI and VLTI/AMBER observations assuming Keplerian rotation \citep{Millour2011}.

Based on the analysis of the observed double-peaked emissions in [\ion{O}{i}], [\ion{Ca}{ii}], and H$\alpha$ lines, \citet{Aret2016} found that the high-density gaseous disk is detached from the star.
This gaseous structure is surrounded by a CO and SiO molecular emitting disk, also in Keplerian rotation, located in turn inside the dusty extended disk \citep{Maravelias2018}.

Ranges for the stellar mass ($9-19$~M$_{\odot}$) and radius ($55-82$~R$_{\odot}$) were derived by \citet{Rovero1994, Snow1994, Hohle2010, Meilland2010, Tetzlaff2011}.
The projected rotational velocity is around $70-80$~km\,s$^{-1}$ \citep{Levato1972, SreedharRao1991, Snow1994}. \citet{Tetzlaff2011} calculated an age of $8.3\pm0.1$~Myr for this star.

\bigskip
\noindent {\bf \object{HK\,Ori}}:
This object is associated with a nebulosity \citep{Bidelman1954} and it is classified as a pre-main sequence Herbig Ae/Be star \citep{Herbig1960} or a HAeB[e] star \citep{Lamers1998}.

\citet{Strom1983} found a variation of the spectral type from the blue to the red part of the spectrum and suggested a combined spectrum of A+F stars.
Later, it was resolved as a binary system with near-IR speckle interferometry by \citet{Leinert1997}, who reported the presence of a second component of T Tauri type.
Furthermore, according to \citet{Baines2004}, the spectral, photometric and polarimetric properties as well as the variability observed in the system indicate that the active object is a G-type T Tauri star with UX Ori characteristics.

\citet{Hillenbrand1992} and \citet{Fairlamb2015} calculated for \object{HK\,Ori} a stellar mass of $2$~M$_{\odot}$ and $1.7$~M$_{\odot}$, respectively.
The former also derived a radius of 1.7 R$_\sun$.
\citet{Smith2005} suggested that the primary component (A) has a circumstellar envelope and derived the parameters for the system: $R_{\rm A} = 1.55$~R$_{\odot}$, $M_{\rm A} = 2$~M$_{\odot}$, $T_{\rm eff_{\rm A}} = 8500$~K, and $R_{\rm B} = 4.1$~R$_{\odot}$, $M_{\rm B} = 1$~M$_{\odot}$,  $T_{\rm eff_{\rm B}} = 4000$~K.
For the circumstellar envelope surrounding the primary component, they calculated a disk inner radius of $31~R_{\odot}$ and an outer radius of $30$~AU.

The age of this star was determined by \citet[][$7.5$~Myr]{Testi1998} and \citet[][$8.73$~Myr]{Fairlamb2015}.

\bigskip
\noindent {\bf  \object{CD$-$24\,5721}}:
\citet{Merrill1950} were the first to report the H$\alpha$ line in emission.
The star presents a large IR excess \citep{Allen1974},  H lines with P~Cygni profiles, and forbidden emission lines \citep{Allen1976}.
It was classified as an unclB[e] star by \citet{Lamers1998}.
            
\citet{Miroshnichenko2003} have suggested that it is a binary system that belongs to a group of Be stars with warm dust \citep[nowadays called FS~CMa stars,][]{Miroshnichenko2007}.
These authors derived a distance to the star of $3.5$~kpc and estimated that it rotates at a projected rotation velocity of $200$~km\,s$^{-1}$.
\citet{Lee2016} also suggested that it is an evolved star that belongs to the FS~CMa group.

\bigskip
\noindent {\bf \object{HD\,85567}}:
This star has been widely studied, however, its evolutionary state is still under discussion.
Some authors suggested that the star is a pre-main sequence Herbig Ae/Be \citep{Oudmaijer1992, The1994} while others considered it as a B[e] interacting binary \citep{Miroshnichenko2001}.

Tables~\ref{StellarParamLit} and \ref{StellarParamLit2} show the stellar parameters found in the literature.
The spectral classification ranges from B2 to B9 and the star has a low rotational speed, $v\, \sin\,i= 31-50$~km\,s$^{-1}$ \citep{Khokhlov2017, Corporon1999, Miroshnichenko2001}.
The stellar mass determinations are: larger than $6$~M$_{\odot}$ \citep{Manoj2006},  $12\pm2$~M$_{\odot}$ \citep[][and a stellar radius of $9\pm2$~R$_{\odot}$]{Verhoeff2012}, and $6^{+2.7}_{-0.18}$~M$_{\odot}$ \citep[][and a stellar radius of $7.2^{+1.5}_{-1.2}$~R$_{\odot}$]{Fairlamb2015}.
In addition, there are two age determinations, $t<0.01$~Myr and $t=0.27^{+0.52}_{-0.18}$~Myr \citep[][respectively]{Manoj2006, Fairlamb2015}. 

\citet{Malfait1998} proposed that the star has a double dust disk, and \citet{Wheelwright2013} concluded that it is a YSO (young stellar object) with an optically thick gaseous disk within a larger dust disk that is being photo-evaporated from the outer edge.
The inner ring-fit radius has $0.8-1.6$~AU \citep{Vural2014}.
Additional temperature-gradient modelling resulted in an extended disk with an inner radius of $0.67^{+0.51}_{-0.21}$~AU, a high inner temperature of $2\,200^{+750}_{-350}$~K, and a disk inclination of ${{53^{\circ}}^{+15}_{-11}}$.
In contrast, the SED analysis performed by \citet{Lee2016} reveals an inner envelope with a size smaller than $1$~AU, at a temperature of $1\,600$~K, and an outer envelope edge of $15$~AU at $300$~K. \citet{Khokhlov2017} argued that the circumstellar gas and dust were produced during the object's evolution more likely as a binary system (containing an undetected secondary component) than as a product of a merger.

\bigskip
\noindent {\bf \object{CPD$-$59\,2854}}:
The star is located at the southern tip of the \object{G289.0–0.3} cloud complex, a prominent peripheral cloud in the field of view of the Carina Nebula region.
$Herschel$ maps reveal \object{CPD$-$59\,2854} as a bright compact far-IR source surrounded by dust, with temperatures of about 18~K and 25~K \citep{Roccatagliata2013}.

The star was included in many catalogues of emission-line stars \citep{Wray1966, Stephenson1971, Jaschek1982} and stars with a strong IR excess \citep{Dong1991}.
\citet{Graham1970} classified it with a spectral type B3IIIe, and derived a $M_V = -2.9$ mag and $d=2.2$ kpc, while \citet{Bailer-Jones2018} placed it at $3.825^{+0.480}_{-0.028}$~kpc.
\citet{Vijapurkar1993} classified this star as a B3Iab star, \citet{The1994} considered it into the group of extreme emission-line objects, and \citet{deWinter2001} listed it as a B[e] star.

\bigskip
\noindent {\bf \object{MWC\,877}}:
The star is listed in many catalogues and surveys of emission-line stars \citep[EM, i.e.,][among others]{Merrill1949, Wray1966, Henize1976, Drilling1981, Jaschek1982, Kozok1985, Vijapurkar1993, The1994, Dong1991}.
According to the intensity of the optical absorption lines of \ion{He}{i} and single ionized atoms, \citet{Carmona2010} classified this star with the spectral type B4~IIe.
Stellar parameters found in the literature are given in Table~\ref{StellarParamLit2}.

\bigskip
\noindent {\bf \object{HD\,323\,771}}:
This star was classified as a Bep shell star by \citet{Herbst1975}.
It is associated with a  moderated surface brightening nebula of type II \citep{vandenBergh1975}.
From the optical photometry, \citet{Kozok1985} derived an interstellar colour excess of $E(B-V)=0.4$ and a distance $d=2$~kpc which is twice the value obtained by \citet{Bailer-Jones2018}.
            
\citet{Merrill1950} were the first to observe a bright  H$\alpha$ line of mid-intensity.
The spectrum shows H lines with P~Cygni profiles and \ion{Fe}{ii} emission lines \citep{Herbst1975}.
\citet{Persi1991} reported the presence of a stellar wind with a velocity of $250-350$~km~s$^{-1}$ and an upper limit for the mass-loss rate of $2 \times 10^{-10}$~M$_{\odot}$~yr$^{-1}$.
These authors also estimated a temperature of $1\,500$~K for the extended dust circumstellar envelope.

Both \citet{Persi1991} and \citet{GregorioHetem1992} identified \object{HD\,323771} as a pre-main sequence Herbig Ae star while \citet{Dufton2001} remarked that the red spectrum has planetary nebulae characteristics.
        
\bigskip
\noindent {\bf \object{HD\,52721}}: 
This star is one of the brightest and most massive objects in the CMa R1 association.
According to \citet{Johnson1955}, this is the central star of S~169 faint nebulae.
\citet{vandenBergh1966} noted  reflection nebulosity and emission from a faint circular disk.
\citet{Finkenzeller1984} remarked that the reflection nebula is located in a region of general obscuration as required for Herbig Be stars.
However, \object{HD\,52721} has IR colours and a (weak) IR-excess as a normal Be star (CBe), which makes its membership in the Herbig Be stars doubtful.

The star is known as a visual binary \citep[\object{ADS\,5713},][]{Aitken1932} with a separation of $1\farcs6$ and a magnitude difference of $\sim 0.7$~mag.
It is also an eclipsing system with an orbital period of $1.61$~days \citep{Pogodin2011}.
\citet{Shokry2018} concluded that the system is a semidetached one, with a secondary evolved from the main sequence.
More recently, via speckle interferometry, it was possible to identify the star as a quadruple system \citep{Obolentseva2021}. 

The star was classified as an emission-line star of spectral type  B2Vne \citep{Guetter1968, Whittet1980},  B3e \citep{Merrill1933}, B1~Vne \citep{Herbst1982}, and as an early-type star with circumstellar shell  \citep{Skiff2014}.
The spectrum shows broad hydrogen absorption lines but those of \ion{He}{i} are fairly narrow and variable in strength \citep{Copeland1963}.
The H$\beta$ line is extremely broad with a variable emission at its centre \citep{Claria1974, Herbst1978}. 

Recently, \citet{Zasche2020}  roughly estimated the angular distance of the predicted third component ($\sim 40$~mas) and they found that the eclipse timing variation (ETV) modulation has a period of about $19.7$ years.

\bigskip
\noindent {\bf \object{HD\,53367}}:
This star is situated at the border of the constellation of Monoceros and is immersed in the Seagull Nebulae \object{IC\,2177}.
It has infrared excess indicative of circumstellar shells \citep{Herbst1978} and shows signs of cool dust in the form of a far-IR excess \citep{Tjin2001}.
The circumstellar material produces a reddening $E(B-V) \sim 0.55$ along the line of sight but without any contribution to the diffuse band strengths. 

The star was included in the young Herbig Ae/Be star list as a B0\,IV:e by \citet{Herbig1960}.
It displays emission lines in the visual spectrum, including H$\alpha$ as the brightest one \citep{Pogodin2006} and shows long-term cyclic photometric variability ($P\sim 9$~years).
In addition, the star is part of a triple star system, showing significant radial velocity variations with a $P\sim 166$ days and an orbit’s eccentricity $e \sim 0.18$ \citep{Corporon1999} or $P \sim 183$ days and $e \sim 0.28$ \citep{Pogodin2006}.
The system consists of a massive ($\sim 20$ M$_\odot$) primary B0e star in the main sequence (if not later) stage and a pre-main sequence secondary object of $M\sim 4-5$~M$_\sun$ with a mean distance of about $1.7$ AU between them. 

\bigskip
\noindent {\bf \object{HD\,58647}}:
The star was reported as an early line-emission star by \citet{MacConnell1981} and then classified as a B9~IV \citep{Houk1988} Herbig star with large IR excess \citep{Oudmaijer1992}. \citet{Guzman-Diaz2021} and \citet{Marcos-Arenal2021} derived the stellar parameters (see Table \ref{StellarParamLit}) and the interferometric size, respectively.
A longitudinal magnetic field of 209 G was also detected \citep{Jarvinen2019}.
The interferometric data and the observed Br$\gamma$ profile can be well reproduced with a disc wind model with its inner radius located just outside of a small magnetosphere \citep{Kurosawa2016}.

\bigskip
\noindent {\bf  \object{LHA\,120$-$S\,65}:}
This star has been scarcely studied.
It was included in the list of H$\alpha$ emission stars in the Large Magellanic Cloud by \citet{Henize1956} and \citet{Andrews1964}.

\onecolumn
\input{tablas/Table_StellarParamLit.tex}

\input{tablas/Table_StellarParamLit2.tex}

\listofobjects

\end{document}

%% file: tablas/Table_obslog.tex
\begin{tabular}{lrccrr}

\hline
\hline \noalign{\smallskip}
 ID & $m_{\rm v}$ & Obs. date  & Observatory & $T_{\rm exp}~~~~~$ & ~~~~~~~~ Reported\\
      &           & yy-mm-dd  &               &    [s]~~~~~~ & classification\\
\hline \noalign{\smallskip}
\object{Hen\,3$-$847}      & $10.58$ & 2012-03-01    & LNA    & $1200$ (1) & HAeBe/post-AGB/sgOB/unclB[e]\\
                           &         & 2001-03-09    & CASLEO & $900/600$ (2) & \\
\object{AS\,202}\tablefootmark{*} & $ 9.72$ & 2012-04-11    & LNA    & $900/1500$ (3) & HAeBe/FS~CMa-type\\
                           &         & 2012-02-05    & CASLEO & $ 900$ (1) & \\
\object{Hen\,3$-$1398}     & $10.62$ & 2011-05-13    & CASLEO & $2000$ (1) & HAeBe/B[e]/nova\\
\object{HD\,62623}\tablefootmark{*} & $ 3.93$ & 2012-02-04    & CASLEO & $  10$ (1) & sgA[e]/sgB[e]\\
\object{HK\,Ori}\tablefootmark{*}     & $11.10$ & 2001-03-09    & CASLEO & $2100$ (1) & HAeBe/HAeB[e]\\
\object{CD$-$24\,5721}\tablefootmark{*} & $10.98$ & 2001-03-08    & CASLEO & $1800$ (1) & unclB[e]/FS~CMa-type\\
\object{HD\,85567}\tablefootmark{*}   & $ 8.57$ & 2001-03-08/10 & CASLEO & $300$ (2) & HAeBe/symB[e]/YSO\\
\object{CPD$-$59\,2854}    & $10.50$ & 2001-03-08    & CASLEO & $1500$ (1) & B[e]\\
\object{MWC\,877}          & $ 9.27$ & 2001-03-10    & CASLEO & $300/600$ (2) & EM\\
\object{HD\,323771}      & $11.23$ & 2001-03-08    & CASLEO & $2200$ (1) & HAeBe\\
\object{HD\,52721}\tablefootmark{*}       & $ 6.59$ & 2012-02-04    & CASLEO & $  60$ (1) & HAeBe/CBe\\
\object{HD\,53367}\tablefootmark{*}       & $ 6.96$ & 2012-02-06    & CASLEO & $60/120$ (2) & HAeBe\\
\object{HD\,58647}       & $ 6.85$ & 2012-02-06    & CASLEO & $120$ (1) & HAeBe\\
\object{LHA\,120$-$S\,65}\tablefootmark{LMC}  & $13.39$ & 2006-01-20    & CASLEO & $ 900$ (1) & EM\\   
\hline
\end{tabular}
\tablefoot{
\tablefoottext{*}{Stars in binary or multiple systems;}
\tablefoottext{LMC}{Star in the Large Magellanic Cloud;}
HAeBe: Herbig Ae/Be star;
sgOB: supergiant OB-type star;
YSO: young stellar object;
EM: emission-line star;
CBe: classical Be star.
}

%% file: tablas/Table_BCDpar.tex
\begin{tabular}{lccccc}
\hline\hline
\noalign{\smallskip}
$ID$          &  $D$  & $\lambda_{\rm 1} - 3700 $  &   $\Phi_{\rm bb}$ &   $\Phi_{\rm b}$ & SBD\\
              &  [dex]     &     [\AA]               & [$\mu$m]    & [$\mu$m]    &\\   
\noalign{\smallskip} \hline \noalign{\smallskip}
\object{Hen\,3$-$847}$^{\rm(LNA)}$   & $0.28$ & $47$   & $\cdots$ & $1.39$   & E\\
\object{Hen\,3$-$847}           & $0.29$ & $43$   & $1.24$   & $\cdots$ & E\\
\object{AS\,202}$^{\rm(LNA)}$     & $0.45$ & $66$   & $\cdots$ & $1.34$   & N\\
\object{AS\,202}                & $0.41$ & $72$   & $\cdots$ & $1.20$   & N\\
\object{Hen\,3$-$1398}          & $0.02$ & $26$   & $1.72$   & $\cdots$ & E\\
\object{HD\,62623}            & $0.30$ & $18$   & $\cdots$ & $1.39$   & A\\
\object{HK\,Ori}                & $0.38$ & $55$   & $1.73$   & $\cdots$ & N\\
\object{CD$-$24\,5721}          & $0.05$ & $61$   & $1.65$   & $\cdots$ & E\\
\object{HD\,85567}            & $0.19$ & $47$   & $1.20$   & $\cdots$ & E\\
\object{CPD$-$59\,2854}         & $0.13$ & $39$   & $1.58$   & $\cdots$ & N\\
\object{MWC\,877}               & $0.04$ & $19$   & $2.86$   & $\cdots$ & E\\
\object{HD\,323\,771}           & $0.23$ & $71$   & $1.47$   & $\cdots$ & N\\
\object{HD\,52721}            & $0.14$ & $82$   & $\cdots$ & $0.97$   & N\\
\object{HD\,53367}            & $0.06$ & $48$   & $\cdots$ & $1.46$   & E\\
\object{HD\,58647}            & $0.41$ & $45$   & $\cdots$ & $0.86$   & A\\
\object{LHA\,120$-$S\,65}       & $0.09$ & $36$     & $0.78$   & $\cdots$ & E\\
\noalign{\smallskip} \hline \noalign{\smallskip}
\end{tabular}

%% file: tablas/Table_ParFun.tex
\begin{tabular}{llllcllc}
\hline
\hline
\noalign{\smallskip}
$ID$ & $ST$ & $T_{\rm eff}$ & $\log g $ & $\Phi^0$ & $M_{\rm v}$~~~ & $M_{\rm bol}$~~~& $A_{\rm v}$ \\
 & & [K] & & [$\mu$m] & [mag]~~ & [mag]~~ & [mag]~~ \\
\noalign{\smallskip} \hline \noalign{\smallskip}
\object{Hen\,3$-$847}$^{\rm(LNA)}$    &  B6\,IV      & $15\,133\pm1234$ & $3.8\pm0.10$    & $0.78\pm0.01$ & $-1.00\pm0.22$ & $-2.00\pm0.32$ & $1.30\pm0.03$\\
\object{Hen\,3$-$847}              &  B6\,IV      & $14\,905\pm1071$ & $3.6\pm0.20$    & $0.78\pm0.01$ & $-1.25\pm0.21$ & $-2.17\pm0.27$ & $1.12\pm0.03$\\
\object{AS\,202}$^{\rm(LNA)}$         &  A1\,V:      & $10\,408\pm 548$ & $4.3\pm0.06$    & $0.98\pm0.03$ & $ 0.75\pm0.18$ & $-0.14\pm0.27$ & $0.78\pm0.06$\\
\object{AS\,202}                   &  A0\,V:      & $10\,335\pm 757$ & $4.5\pm0.06$    & $0.93\pm0.03$ & $ 0.65\pm0.39$\tablefootmark{C} & $0.33\pm0.42$\tablefootmark{F} & $0.56\pm0.06$\\
\object{Hen\,3$-$1398}        &  B0-1\,Ia      &  $25\,000\pm1500$ & $2.9\pm0.11$\tablefootmark{H} & $0.69\pm0.01$ & $-6.43\pm0.03$\tablefootmark{C} & $-8.62\pm0.15$\tablefootmark{F} & $2.39\pm0.03$\\
\object{HD\,62623}          &  B9\,Ib      & $11\,136\pm 318$ & $1.5\pm0.05$\tablefootmark{H} & $0.88\pm0.02$ & $-6.25\pm0.05$\tablefootmark{C} & $-6.75\pm0.09$\tablefootmark{F} & $1.09\pm0.03$\\
\object{HK\,Ori}              &  B9\,V       & $11\,940\pm 380$ & $4.1\pm0.06$    & $0.85\pm0.02$ & $ 0.06\pm0.24$ & $-0.78\pm0.14$ & $2.05\pm0.06$\\
\object{CD$-$24\,5721}        &  O8\,IV      & $36\,667\pm3503$ & $3.94\pm0.09$   & $0.66\pm0.02$ & $-4.50\pm0.77$ & $-7.00\pm0.39$ & $2.29\pm0.06$\\
\object{HD\,85567}          &  B3\,III     & $20\,000\pm1639$ & $3.7\pm0.16$    & $0.73\pm0.01$ & $-2.33\pm0.33$ & $-3.83\pm0.42$ & $1.12\pm0.03$\\
\object{CPD$-$59\,2854}       &  B2\,II      & $21\,000\pm 917$ & $2.9\pm0.08$    & $0.71\pm0.01$ & $-4.25\pm0.47$ & $-5.50\pm0.50$ & $2.02\pm0.03$\\
\object{MWC\,877}             &  B3\,Ia:     & $20\,000\pm2292$ & $2.5\pm0.20$\tablefootmark{H} & $0.72\pm0.01$ & $-6.33\pm0.07$\tablefootmark{C} & $-8.22\pm0.26$\tablefootmark{F} & $4.96\pm0.03$\\
\object{HD\,323\,771}         &  B5\,V:      & $17\,500\pm1250$ & $4.3\pm0.02$   & $0.76\pm0.01$  & $-1.20\pm0.19$ & $-2.46\pm0.34$ & $1.80\pm0.03$\\
\object{HD\,52721}          &  B2\,V:     & $27\,361\pm2611$:& $4.4\pm0.09$:  & $0.74\pm0.03$: & $-1.61\pm0.47$: & $-3.26\pm1.09$: & $0.50\pm0.06$\\
\object{HD\,53367}          &  B0\,III     & $32\,113\pm4616$ & $2.9\pm0.16$   & $0.68\pm0.01$ & $-4.6\pm1.35$    & $-6.77\pm0.58$  & $1.30\pm0.03$\\
\object{HD\,58647}          &  B9\,IV      & $12\,203\pm 600$ & $3.4\pm0.14$   & $0.87\pm0.02$ & $-0.50\pm0.13$ & $-1.25\pm0.19$ & $0.00\pm0.06$\\
\object{LHA\,120$-$S\,65}     &  B2\,Ib                         & $22\,000\pm1031$ & $2.6\pm0.08$\tablefootmark{H} & $0.70\pm0.01$  & $-5.67\pm0.63$                  & $-7.00\pm0.62$                  & $0.19\pm0.03$\\
\hline
\end{tabular}
\tablefoot{
  $\Phi^0$ is $\Phi_{b}^0$ or $\Phi_{bb}^0$ according to the wavelength range.
  The symbol ":" indicates extrapolated values.
  \tablefoottext{H}{The $\log g$ values are estimated from the $T_{\rm{eff}}-\log\,g$ relationship ($\log g = 3.98 \log T_{\rm eff} - 14.64$) found by \citet{Haucke2018}.}
  \tablefoottext{C}{$M_{\rm v}$ values are estimated from tables given by \citet{Cox2000}.}
  \tablefoottext{F}{$M_{\rm bol}$ values were calculated using the bolometric correction given by \citet{Flower1996}.}
}

%% file: tablas/Table_ParFun2.tex
\begin{tabular}{lccccc}
\hline
\hline
\noalign{\smallskip}
~~~~ID     &      $\log (L/L_{\sun})$ &  $\log (\mathcal{L}/\mathcal{L}_{\sun})$\tablefootmark{*} & 
$ (m-M_{\mathrm{v}})_0$ & $d$ & $~~d_{G}$\tablefootmark{**}\\
           &   &                      & [mag] & [pc] & [pc]\\ 
\hline \noalign{\smallskip}
\object{Hen\,3$-$847}$^{\rm(LNA)}$   & $2.69\pm0.13$ & $2.31\pm0.10$ & $10.26\pm0.23$ & $1127\pm118$  & $649_{-63}^{+93}$\\
\object{Hen\,3$-$847}$^{\rm(CASLEO)}$ & $2.76\pm0.11$ & $2.48\pm0.20$ & $10.79\pm0.22$ & $1440\pm143$  & $649_{-63}^{+93}$\\
\object{AS\,202}$^{\rm(LNA)}$        & $1.94\pm0.11$ & $1.16\pm0.06$ & $ 8.20\pm0.20$ & $ 437\pm 40$  & $335_{-4}^{+3}$\\
\object{AS\,202}$^{\rm(CASLEO)}$      & $1.76\pm0.17$ & $0.95\pm0.06$ & $ 8.50\pm0.40$ & $ 502\pm 91$  & $335_{-4}^{+3}$\\
\object{Hen\,3$-$1398}             & $5.34\pm0.06$ & $4.11\pm0.11$ & $14.66\pm0.06$ & $8536\pm236$  & $1780_{-57}^{+83}$\\
\object{HD\,62623}               & $4.59\pm0.04$ & $4.11\pm0.05$ & $ 9.11\pm0.18$ & $ 662\pm 54$  & $1124_{-198}^{+357}$\\
\object{HK\,Ori}                   & $2.20\pm0.06$ & $1.56\pm0.06$ & $ 9.00\pm0.24$ & $ 630\pm 71$  & $\cdots$\\
\object{CD$-$24\,5721}             & $4.69\pm0.16$ & $3.71\pm0.10$ & $13.14\pm0.77$ & $4245\pm1515$ & $3255_{-164}^{+148}$\\
\object{HD\,85567}               & $3.42\pm0.17$ & $2.99\pm0.16$ & $ 9.80\pm0.33$ & $ 913\pm138$  & $1036_{-15}^{+17}$\\
\object{CPD$-$59\,2854}            & $4.09\pm0.20$ & $3.82\pm0.08$ & $12.72\pm0.48$ & $3505\pm768$  & $3144_{-108}^{+104}$\\
\object{MWC\,877}                  & $5.18\pm0.10$ & $4.12\pm0.21$ & $10.57\pm0.08$ & $1297\pm 46$  & $1501_{-56}^{+69}$\\
\object{HD\,323\,771}              & $2.87\pm0.14$ & $2.02\pm0.04$ & $10.78\pm0.20$ & $1432\pm129$  & $1074_{-29}^{+26}$\\
\object{HD\,52721}               & $3.19\pm0.44$ & $2.72\pm0.10$ & $7.72\pm0.42$  & $479\pm 76$\tablefootmark{***}  & $\cdots$\\
\object{HD\,53367}               & $4.6\pm0.23$ & $4.49\pm0.17$ & $10.27\pm1.35$ & $1131\pm701$  & $1599_{-458}^{+926}$\\
\object{HD\,58647}               & $2.39\pm0.08$ & $2.32\pm0.14$ & $ 7.35\pm0.13$ & $295\pm7.67$  & $302_{-2}^{+2}$\\
\object{LHA\,120$-$S\,65}      & $4.69\pm0.25$ & $4.12\pm0.08$ & $18.87\pm0.63$ & $59564\pm17305$ & $27343_{-6129}^{+13324}$\\
\hline
\end{tabular}
\tablefoot{
  \tablefoottext{*}{$\log (\mathcal{L}/\mathcal{L}_{\sun}) = 4  \log(T_{\rm eff}) - \log\,g - \log \mathcal{L}_{\sun}$ \citep{Langer2014}.}
  \tablefoottext{**}{Distance values taken from \citet{Bailer-Jones2021}.
  \tablefoottext{***}{The distance was calculated assuming a binary system with a difference of $0.7$~mag in $m_{\rm v}$ \citep{Aitken1932}.}
 }}

%% file: tablas/Table_StellarParamLit.tex
\begin{longtable}{llcccc}
\caption{Stellar parameters for our sample of stars taken from the literature. The numbers in brackets give the references related to the reported measurements.\label{StellarParamLit}}\\
\hline\hline
$ID$          &  $ST$  &  $T_{\rm eff}$ & $\log g$ & $M_{\rm bol}$ & $M_{\rm v}$\\
              &        &      [K]     &          & [mag]       & [mag]\\   
\hline

\endfirsthead

\multicolumn{6}{l}%
{\tablename\ \thetable\ -- \textit{Continued from previous page.}} \\
\hline\hline
$ID$          &  ~~~~~$ST$  & ~~~~~~~~ $T_{\rm eff}$ & ~~~$\log g$ & ~~~~~~~~$M_{\rm bol}$ & ~~~~~~$M_{\rm v}$\\
              &        &     ~~~~~~~~ [K]     & ~~  [dex]       & ~~~~~~~~[mag]       & ~~~~~~[mag]\\   
\hline
\endhead
\hline \multicolumn{6}{r}{\textit{Continued on next page.}} \\
\endfoot

\object{Hen\,3$-$847}      & B5e (39, 48)           & $\cdots$   &   $\cdots$     & $\cdots$             & $\cdots$           \\
   & B6/A0:[e] (41)     &     $\cdots$    &   $\cdots$     &   $\cdots$                   &         $\cdots$           \\
                           &  B6:III[e] (43)      & $14\,125$ (43)         & $\cdots$                  &      $\cdots$                &   $\cdots$                 \\
                                & $\cdots$                        &      $15\,100\pm1100$ (61)              &   $3.84$ (61)   & $\cdots$                                  &  $\cdots$                  \\
                                   &  $\cdots$    & $14\,000$ (63)         & $3.80$ (63)        &   $\cdots$                   &         $\cdots$           \\
                           & B6 (64)           &  $\cdots$                    &   $\cdots$                &    $\cdots$                  & $\cdots$                   \\
\hline
\object{AS\,202}           &   B9III-IV (20)          & $\cdots$             & $\cdots$          & $\cdots$             & $-0.75$ (20)       \\
                           & A+K (54)     & $\cdots$                     &    $\cdots$               &       $\cdots$               & $\cdots$                   \\
                           & A0-2 (67) & $9\,500\pm500$ (67)   &  $\cdots$  &  $2.73\pm0.13$ (67)& $\cdots$\\
\hline
\object{Hen\,3$-$1398}     &  OB+CE (12)          &    $\cdots$                    &       $\cdots$              &     $\cdots$                  &   $\cdots$                   \\
  &   Oe (39)        &       $\cdots$                 &               $\cdots$      &    $\cdots$    &       $\cdots$               \\
                           & O9/B0 (45)   & $30\,000$ (45)  & $4.0$ (45)        & $\cdots$             & $\cdots$           \\
                           &  B1Ve (57)          & $25\,400$ (57)         &     $\cdots$ &      $\cdots$       &       $\cdots$               \\
\hline
\object{HD\,62623}       & A2p (1)            & $\cdots$    & $\cdots$  & $\cdots$           & $\cdots$         \\
                           & cA2ep (3)          & $\cdots$          & $\cdots$       & $\cdots$                &  $\cdots$                  \\
                           & A3IIIp (4)         &  $\cdots$           &   $\cdots$                & $\cdots$                     &  $\cdots$       \\
                           & A2Iab (5, 10, 13)          & $\cdots$        & $\cdots$       &  $\cdots$                &   $\cdots$                 \\
                           & A3IIep (6, 25)         & $\cdots$          &  $\cdots$                 &  $\cdots$                    &    $\cdots$                \\
                           & B8Iab: (12)        &  $\cdots$                    &    $\cdots$               &  $\cdots$                    & $\cdots$                   \\
                           & A2Ib (19, 31)          &     $\cdots$                 & $\cdots$                  &  $\cdots$                   &        $\cdots$            \\
                           & A2Iabe (24, 30, 35, 37)        &   $9\,080$ (37, 59)                  & $\cdots$                  &     $-6.8$ (37)                 &  $-6.5$ (37)             \\
                            & A3IIpe (28)        & $\cdots$                     &  $\cdots$                 &  $\cdots$                  &     $\cdots$               \\
                           & A2I (32)           &$\cdots$                       &  $\cdots$                  &  $\cdots$                     & $\cdots$                    \\
                           & A3Iab (36)         & $8\,500$ (36)          & $2.0$ (36)                  & $\cdots$                      &  $-5.3$ (36)                  \\
                               & $\cdots$          & $9\,000$ (40)          & $1.35$ (40)       &  $\cdots$                &   $\cdots$                 \\
                           & A2.7Ib (58)        &  $\cdots$                     & $\cdots$                   &  $\cdots$                     &  $-5.5\pm0.3$ (58)                   \\
                                &  $\cdots$           & $8\,250\pm250$ (60)    & $2.0\pm0.5$ (60)  & $\cdots$           & $\cdots$         \\
\hline
\object{HK\,Ori}                   & Aep (3)              &       $\cdots$                 &       $\cdots$              &         $\cdots$               &     $\cdots$                 \\
                           & A4ep (4)               &       $\cdots$                 &       $\cdots$              &         $\cdots$               &     $\cdots$                 \\

 & B7-8/A4 (21)           & $12\,500$ (21)         & $3.30$ (21)        & $\cdots$             & $\cdots$           \\
                &  B8[e] (26)            &     $\cdots$                   &        $\cdots$             &     $\cdots$                   &     $\cdots$                 \\
                                      & A4 (27)  &   $\cdots$                     &        $\cdots$             &          $\cdots$              &    $\cdots$                  \\
                           & A5 (33)         & $8\,318$ (33)          &  $\cdots$     &      $\cdots$                 &      $\cdots$               \\
                           & A9V (38)           &     $\cdots$                   &    $\cdots$                 &    $\cdots$                    &          $\cdots$            \\
                           &    $\cdots$             & $8\,511$ (42)          &      $\cdots$               &      $\cdots$                  &       $\cdots$               \\
                           & A+G (50)           &     $\cdots$                   &        $\cdots$             &     $\cdots$                   &     $\cdots$                 \\
                           &  $\cdots$      & $8\,500+4000$ (52)     &    $\cdots$                 &    $\cdots$                    &       $\cdots$               \\
                           &       $\cdots$        & $8500$ (63)          &       $4.22$ (63)              &     $\cdots$                   &   $\cdots$                   \\
\newpage
\hline
\object{CD$-$24\,5721}     & B1.5V: (44)             & $25800\pm1200$ (44)  & $4.4$ (44)        & $-5.28\pm0.32$ (44)  & $\cdots$           \\
                           & Oe (16)        &  $\cdots$                     &        $\cdots$            &   $\cdots$                    &   $\cdots$                  \\
\hline
\object{HD\,85567}       & B5Vne (15)         &  $\cdots$      & $\cdots$       & $\cdots$             & $\cdots$           \\
& B8 (17)        & $\cdots$      & $\cdots$       & $\cdots$             & $\cdots$           \\
& B8V:ne (18)         & $\cdots$      & $\cdots$       & $\cdots$             & $\cdots$           \\
&   B7/8Ve (43)      & $12\,590$ (43)         & $\cdots$       & $\cdots$             & $\cdots$           \\
                           &       B2V (45)      & $19\,000$ (45)          & $3.5$ (45)    &    $\cdots$                  & $\cdots$                   \\
                           & B9 (48)     & $\cdots$      & $\cdots$       & $\cdots$             & $\cdots$           \\
                           & B8Vne (51)         &                 $\cdots$     &  $\cdots$                 &           $\cdots$           &  $\cdots$                  \\
                           &  $\cdots$      & $21\,880$ (53)         & $\cdots$  &  $\cdots$                    &  $\cdots$                  \\
                           &   $\cdots$      & $12\,450$ (56)         &          $\cdots$         &   $\cdots$                   &     $\cdots$               \\
                           &     $\cdots$       & $13\,000\pm500$ (63)    &  $3.5\pm0.3$ (63)                & $\cdots$                     &  $\cdots$                  \\
                       &    $\cdots$         & $15\,000\pm500$ (65)  & $\sim 4.0$ (65)                   &  $\cdots$                    &  $\cdots$                  \\
                           
\hline
\object{CPD$-$59\,2854}    & B3IIIe (11)        & $\cdots$             & $\cdots$          & $\cdots$             & $-2.9$ (11)        \\
                           & B3Iab (34)         & $\cdots$                     &    $\cdots$               &  $\cdots$                    &  $\cdots$                  \\
                           & B2/B3Ve (39)       &           $\cdots$           &    $\cdots$               &  $\cdots$                    &    $\cdots$                \\
\hline
\object{MWC\,877}          & BIe (29)           &   $\cdots$    & $\cdots$          & $\cdots$             & $-6.45$ (29)       \\
                           & B3Iane$_{2+}$ (34) &         $\cdots$             &   $\cdots$                &  $\cdots$                    &  $\cdots$                  \\
                           & B4IIe (57)         &       $15\,000$ (57)                &    $\cdots$               & $\cdots$                     &  $\cdots$                  \\
\hline
\object{HD\,323\,771}    & Bep (14)            &   $\cdots$     & $\cdots$          & $\cdots$             &   $\cdots$      \\
                           & B8 (17)            & $\cdots$                      &  $\cdots$                  &   $\cdots$                  &   $\cdots$                  \\
                                                      &  $\cdots$            &  $\cdots$                     &  $\cdots$                  &  $\cdots$                     &  $-1.63$ (29)                  \\
                           & B5Vp (48)          &        $15\,000$ (48)               &   $\cdots$                 &  $\cdots$                     & $\cdots$                    \\
\hline
\object{HD\,52721} 
&$\cdots$&$22\,500$ (63) &$3.9$ (63) &$\cdots$&$\cdots$\\
&B3e (2) &$\cdots$&$\cdots$&$\cdots$&$\cdots$\\
&B2Vne (9) &$\cdots$&$\cdots$&$\cdots$&$\cdots$\\
&B1Ve (23)&$\cdots$&$\cdots$&$\cdots$& $\cdots$\\
&B2V (22) &$\cdots$&$\cdots$&$\cdots$&$\cdots$\\
& $\cdots$ & $25\,000$ (68) & $4.0$ (68) &$\cdots$&$\cdots$\\
&B2 (7) &$\cdots$&$\cdots$&$\cdots$&$\cdots$\\
\hline
\object{HD\,53367}          &B0IV:e (8)&$\cdots$&$\cdots$&$\cdots$&$\cdots$\\
&$\cdots$&$29\,500$ (63)& $4.25$(63) &$\cdots$&$\cdots$\\
&B0.5: (22, 7)&$\cdots$&$\cdots$&$\cdots$&$\cdots$\\
&$\cdots$&$29\,000\pm2\,000$ (62)& $4.0$ (62) &$\cdots$&$\cdots$\\
&B0IV (47)&$\cdots$&$\cdots$&$\cdots$&$\cdots$\\
\hline
\object{HD\,58647}          
& $\cdots$ & $10\,750\pm125$ (69) & $\cdots$ & $\cdots$&$\cdots$
\\
& B9IV (22) & $\cdots$ & $\cdots$ & $\cdots$& $\cdots$
\\
& $\cdots$ & $10\,500\pm200$ (66,55) & $3.3$ (55) & $\cdots$&
\\
&A0 IIIn (49) &  $\cdots$& $\cdots$ & $\cdots$&$\cdots$
\\
&B9IVep (46) &  $\cdots$& $\cdots$ & $\cdots$&$\cdots$
\\
\hline
\object{LHA\,120$-$S\,65} & $\cdots$ & $\cdots$ & $\cdots$ & $\cdots$& $\cdots$\\      
\hline
\end{longtable}
\tablebib{
(1)  \citet{Lunt1919};
(2) \citet{Merrill1933};
(3)  \citet{Merrill1943};
(4)  \citet{Bidelman1954};
(5)  \citet{deVaucouleurs1957};
(6)  \citet{Kron1958};
(7) \citet{Mendoza1958};
(8)  \citet{Herbig1960};
(9) \citet{Guetter1968};
(10)  \citet{Hiltner1969};
(11)  \citet{Graham1970};
(12)  \citet{Stephenson1971};
(13)  \citet{Levato1972};
(14)  \citet{Herbst1975};
(15)  \citet{Houk1975};
(16)  \citet{Nordstrom1975};
(17)  \citet{Henize1976};
(18)  \citet{Garrison1977};
(19)  \citet{Cucchiaro1978};
(20)  \citet{Eggen1978};
(21)  \citet{Garrison1978};
(22) \citet{Jaschek1980};
(23) \citet{Herbst1982};
(24)  \citet{Hoffleit1982};
(25)  \citet{Fernie1983};
(26)  \citet{Whittet1983};
(27)  \citet{Finkenzeller1984};
(28)  \citet{Reed1984};
(29)  \citet{Kozok1985};
(30)  \citet{Gutierrez-Moreno1986};
(31)  \citet{Gray1987};
(32)  \citet{SreedharRao1991};
(33)  \citet{Hillenbrand1992};
(34)  \citet{Vijapurkar1993};
(35)  \citet{Danks1994};
(36)  \citet{Rovero1994};
(37)  \citet{Snow1994};
(38)  \citet{Terranegra1994};
(39)  \citet{The1994};
(40)  \citet{Plets1995};
(41)  \citet{Lamers1998};
(42)  \citet{Testi1998};
(43)  \citet{vandenAncker1998};
(44)  \citet{Cidale2001};
(45)  \citet{Miroshnichenko2001};
(46) \citet{Mora2001};
(47) \citet{Tjin2001};
(48)  \citet{Vieira2003};
(49) \citet{Abt2004};
(50)  \citet{Baines2004};
(51)  \citet{Zhang2004};
(52)  \citet{Smith2005};
(53)  \citet{Manoj2006};
(54)  \citet{Miroshnichenko2007b};
(55) \citet{Montesinos2009};
(56)  \citet{Acke2010};
(57)  \citet{Carmona2010};
(58)  \citet{Chentsov2010};
(59)  \citet{Hohle2010};
(60)  \citet{Meilland2010};
(61)  \citet{Verhoeff2012};
(62) \citet{Alecian2013};
(63)  \citet{Fairlamb2015};
(64) \citet{Ababakr2016};
(65) \citet{Khokhlov2017};
(66) \citet{Vioque2018};
(67) \citet{Condori2019};
(68) \citet{Pogodin2020};
(69) \citet{Guzman-Diaz2021}.
}

%% file: tablas/Table_StellarParamLit2.tex
\begin{longtable}{lccccc}
\caption{\label{StellarParamLit2}
 Stellar parameters for our sample of stars taken from the literature. The numbers in brackets give the references related to the reported measurements}.\\
\hline\hline
$ID$          &  $\log L_{\star}$  & $E(B-V)$ & $d$   & $A_{\rm v}$\\
              &  [$L_{\odot}$]     & [mag]    & [kpc] & [mag]\\   
\hline
\endfirsthead
\multicolumn{5}{l}%
{\tablename\ \thetable\ -- \textit{Continued from previous page.}} \\
\hline\hline
$ID$          &  $\log L_{\star}$  & $E(B-V)$ & $d$   & $A_{\rm v}$\\
              &  [$L_{\odot}$]     & [mag]    & [kpc] & [mag]\\   
\hline
\endhead
\hline \multicolumn{5}{r}{\textit{Continued on next page.}} \\
\endfoot

\object{Hen\,3$-$847}      & $> 0.72$ (15)             &  $\cdots$
        & $> 0.14$ (15)                   & $0.74$ (15)\\
&   $\cdots$         &    $0.18$ (20)                   &          $\cdots$           & $\cdots$\\
                           & $3.22\pm0.08$ (31)        &  $\cdots$                     & $1.66\pm0.23$ (31)              & $0.8\pm0.1$ (31)\\
                           & $2.85$ (32)               &  $\cdots$                     & $1.843$ (32)                    & $0.57$ (32)\\
                           &  $\cdots$                         & $0.16\pm0.02$ (33)   & $2.47\pm0.57$ (33)             &$\cdots$\\
                           &  $\cdots$                         & $\cdots$                      & $0.997_{-0.218}^{+0.379}$ (35) &$\cdots$\\
                           &     $\cdots$                      &   $\cdots$                    & $0.649_{-0.063}^{+0.093}$ (38) &$\cdots$\\
\hline
\object{AS\,202}           & $0.81\pm0.13$ (37)             &      $\cdots$      & $1.5$ (3)                      &$\cdots$\\
                           &  $\cdots$                         &  $\cdots$                     & $0.353_{-0.006}^{+0.006}$ (35) &$\cdots$\\
                           & $\cdots$                          & $\cdots$                      & $0.335_{-0.004}^{+0.003}$ (38) &$\cdots$\\
\hline
\object{Hen\,3$-$1398}     & $5.3\pm0.2$ (18)          & $1.10$ (18)           & $3.3\pm0.4$ (18)                &$\cdots$\\
                           & $\cdots$                          & $0.74\pm0.15$ (26)    & $2.1^{+0.9}_{-0.6}$ (26)        & $2.3\pm0.5$ (26)\\
                           & $4.77\pm0.07$ (31)        & $\cdots$                  & $\cdots$ & $2.4\pm0.1$ (31)\\
                             & $\cdots$                          & $\cdots$   &               $1.958_{-0.190}^{+0.235}$ (35)      &$\cdots$\\
                           & $\cdots$                          & $\cdots$                       & $1.780_{-0.057}^{+0.083}$ (38) &$\cdots$\\
\hline
\object{HD\,62623}       & $\cdots$                & $0.11$ (6)           &  $\cdots$                   & $\cdots$\\                                       &  $\cdots$                         & $\cdots$          & $0.7$ (10, 16, 27)                      &$\cdots$\\
                           &  $\cdots$                         & $0.15$ (11)           & $\cdots$           &$\cdots$\\
                           &  $\cdots$                         & $0.17$ (12)           & $\cdots$                      &$\cdots$\\
                            & $156$ (28)                & $\cdots$           &  $\cdots$                   & $\cdots$\\  
                           &  $\cdots$                         & $\cdots$                      & $0.65\pm0.1$ (29)               &$\cdots$\\
                           & $\cdots$                          &           $\cdots$            & $0.664_{-0.140}^{+0.240}$ (35) &$\cdots$\\
                           &  $\cdots$                         & $\cdots$                      & $1.124_{-0.198}^{+0.357}$ (38) &$\cdots$\\
\hline
\object{HK\,Ori}           & $\cdots$ & $\cdots$             & $0.46$ (4)                     & $1.5$ (4) \\
                           & $\cdots$                          &   $\cdots$                    &    $\cdots$                             & $0.8$ (5) \\
                           & $1.09$ (8)               &    $\cdots$                   & $0.46$ (8)                     & $1.2$ (8) \\
                            & $1.32$ (14)               &    $\cdots$                   & $\cdots$                    & $\cdots$\\
                           & $1.13$ (32)               &    $\cdots$                   & $0.40$ (32)                     & $1.21$ (32)\\
\hline
\object{CD$-$24\,5721}     & $\cdots$         &   $\cdots$       &  $\cdots$               & $1.7$ (17)\\
& $4.4\pm0.2$ (21)          & $\cdots$       & $3.5\pm0.5$ (21)                & $\cdots$\\
& $\cdots$        & $0.73$ (24)           & $\cdots$               & $\cdots$\\
                           &  $\cdots$                         &  $\cdots$                     & $3.612_{-0.442}^{+0.576}$ (35) &$\cdots$\\
                           &  $\cdots$                         &  $\cdots$                     & $3.255_{-0.164}^{+0.148}$ (38) &$\cdots$\\
\hline
\object{HD\,85567}       & $\cdots$              & $0.23$ (13)           & $\cdots$                   & $\cdots$\\
& $>2.54$ (15)              & $\cdots$        & $>0.48$ (15)                    & $0.81$ (15)\\
                           & $4.0\pm0.3$ (18)          & $0.4$ (18)            & $1.5\pm0.5$ (18)                & $\cdots$ \\
                           & $4.66$ (23)               & $0.36$ (23)           &  $\cdots$    & $2.23$ (23) \\
                           & $4.17\pm0.16$ (31)        & $\cdots$                       & $\cdots$                & $1.1\pm0.1$ (31) \\
                           & $3.13^{+0.46}_{-0.45}$ (32) &         $\cdots$             & $0.91^{+0.18}_{-0.15}$ (32)  & $0.89^{+0.03}_{-0.02}$ (32)\\
                           & $3.3\pm0.2$ (34)         &  $\cdots$                      & $1.3\pm0.1$ (34) &$\cdots$ \\
                            &   $\cdots$      & $\cdots$                      & $1.002_{-0.028}^{+0.030}$ (35)  &$\cdots$\\
                           &   $\cdots$      & $\cdots$                      & $1.036_{-0.015}^{+0.017}$ (38) &$\cdots$\\
\hline
\object{CPD$-$59\,2854}    &   $\cdots$               & $\cdots$       & $2.2$ (2)                      & $1.7$ (2) \\
                           &  $\cdots$                         &  $\cdots$                     & $3.825_{-0.386}^{+0.480}$ (35) & $\cdots$\\
                           &  $\cdots$                         & $\cdots$                      & $3.144_{-0.108}^{+0.104}$ (38) & $\cdots$\\
\hline
\object{MWC\,877}          & $\cdots$              & $1.22$ (4)           & $0.54$ (4)                     & $\cdots$\\
                           &  $\cdots$                         & $1.10$ (7)           & $2.56$ (7)                     &$\cdots$\\
                                  &  $\cdots$                        & $1.31$\tablefootmark{*} (9)  &   $\cdots$    &  $\cdots$\\              
                           &  $\cdots$                        & $1.10\pm0.15$ (26) & $1.1^{+0.3}_{-0.2}$ (26) &$3.4\pm0.5$ (26)\\
                           & $\cdots$                          &  $\cdots$   & $1.270_{-0.120}^{+0.147}$ (35) & $\cdots$\\
                           &   $\cdots$                        &  $\cdots$                     & $1.501_{-0.056}^{+0.069}$ (38) & $\cdots$\\
\hline
\object{HD\,323\,771}   
   & $\cdots$                  & $0.40$ (7)           & $2.05$ (7)                     & $\cdots$\\
      & $\cdots$                  & $\cdots$        & $\cdots$                   & $0.9$ (30)\\
                           & $\cdots$                          & $\cdots$                      & $1.072_{-0.061}^{+0.069}$ (35) &$\cdots$\\
                           &   $\cdots$                        & $\cdots$                      & $1.074_{-0.029}^{+0.026}$ (38) &$\cdots$\\
\hline
\object{HD\,52721}       & $3.87$ (32) & $\cdots$& $0.531$ (32) &$\cdots$\\
& $\cdots$ &$\cdots$& $0.72$ (1) &$\cdots$\\ 
\hline
\object{HD\,53367}  
&$4.11$ (32) &$\cdots$& $0.34$ (32) &$\cdots$\\ 
&$\cdots$&$\cdots$&$1.599_{-0.458}^{+0.926}$ (38) &$\cdots$\\
& $\cdots$& 0.55 (19)& $\cdots$&$\cdots$\\

\hline
\object{HD\,58647}       &$2.47$  (22)&$\cdots$& $0.302_{-0.002}^{+0.002}$ (38) &$\cdots$\\
      &$2.49\pm0.01$ (39) &$\cdots$& $0.28^{+0.008}_{-0.005}$ (15)&$\cdots$\\
      &$2.44_{-0.09}^{+0.11}$ &$0.318$ (36) & $\cdots$&$0.37$ (36)\\
      &$2.96$ (25) &$0.543$ (25)& $\cdots$&$\cdots$\\
\hline
\object{LHA\,120$-$S\,65} & $\cdots$ & $\cdots$ & $\cdots$ & $\cdots$\\      
\hline
\end{longtable}
\tablefoot{
\tablefoottext{*}{We {\bf infered} the colour excess $E(B-V)$ from the relation $E(b-y) = 0.74\, E(B-V)$.}
}
\tablebib{
(1) \citet{Mendoza1958};
(2)  \citet{Graham1970};
(3)  \citet{Eggen1978};
(4)  \citet{Finkenzeller1984};
(5)  \citet{Harvey1984};
(6)  \citet{Reed1984};
(7)  \citet{Kozok1985};
(8)  \citet{Hillenbrand1992};
(9)  \citet{Kilkenny1993};
(10)  \citet{Rovero1994};
(11)  \citet{Snow1994};
(12)  \citet{Plets1995};
(13)  \citet{Malfait1998};
(14)  \citet{Testi1998};
(15)  \citet{vandenAncker1998};
(16)  \citet{Bittar2001};
(17)  \citet{Cidale2001};
(18)  \citet{Miroshnichenko2001};
(19) \citet{Tjin2001};
(20)  \citet{Gauba2003};
(21)  \citet{Miroshnichenko2003};
(22) \citet{Monnier2005};
(23)  \citet{Manoj2006};
(24)  \citet{Miroshnichenko2007};
(25) \citet{Montesinos2009};
(26)  \citet{Carmona2010};
(27)  \citet{Chentsov2010};
(28)  \citet{Hohle2010};
(29)  \citet{Meilland2010};
(30)  \citet{Sartori2010};
(31)  \citet{Verhoeff2012};
(32)  \citet{Fairlamb2015};
(33) \citet{Vickers2015};
(34) \citet{Khokhlov2017};
(35) \citet{Bailer-Jones2018};
(36) \citet{Vioque2018};
(37) \citet{Condori2019};
(38) \citet{Bailer-Jones2021};
(39) \citet{Marcos-Arenal2021}.
}